\documentclass{aa}  

\usepackage{graphicx}
\usepackage{txfonts}
\begin{document}

   \title{Kinematic study of planetary nebulae in NGC\,6822\thanks{Based on data obtained at Las Campanas Observatory, Carnegie Institution, Chile.}\thanks{Based on data collected at the Observatorio Astron\'omico Nacional, SPM. B.C., M\'exico.}}

   \author{
   S. N. Flores-Dur\'an\inst{1}, 
  M. Pe\~na\inst{1},
  L. Hern\'andez-Mart\'inez\inst{1},
 J. Garc\'ia-Rojas\inst{2},
 \and M. T. Ruiz\inst{3}}

   \institute{Instituto de Astronom\'\i{}a,
  Universidad Nacional Aut\'onoma de M\'exico,
   Apdo. Postal 70264, M\'exico D.F., 04510, M\'exico\\
              \email{sflores,miriam@astro.unam.mx}
         \and
             Instituto de Astrof{\'\i}sica de Canarias, E-38200, La Laguna, Tenerife, Spain and\\
                    Departamento de Astrof\'{\i}sica, Universidad de La Laguna. E-38205. La Laguna, Spain.\\
                      \email{jogarcia@iac.es}
                     \and
                     Depto. de Astronom{\'\i}a, Universidad de Chile, Casilla 36D, Las Condes, Santiago, Chile\\
                     \email{mtruiz@das.uchile.cl}
             }

   \date{Received 05/06/2013; accepted 13/05/2014}
   
   \titlerunning{Kinematics of PNe in NGC\,6822}

\authorrunning{Flores-Dur\'an et al.} 

 
  \abstract
   {The kinematics of planetary nebulae in external galaxies and in our own is a clue for understanding the behavior of the low-intermediate mass stars and their relation with other components  of the galaxies.}
   {By measuring precise radial velocities of planetary nebulae (which belong to the intermediate-age population), H\,{\sc ii} regions and A-type supergiant stars (which are members of the young population)  in NGC\,6822, we aim to determine whether both types of population share the kinematics of the  disk of H\,{\sc i} found in this galaxy. }
   {Spectroscopic data for six planetary nebulae were obtained with the high spectral-resolution spectrograph Magellan Inamori Kyocera Echelle (MIKE) on the Magellan telescope  at Las Campanas Observatory. Data for another  three PNe and one H\,{\sc ii} region were obtained from the SPM Catalog of Extragalactic Planetary Nebulae, which employed the Manchester Echelle Spectrometer attached to the 2.1m telescope at the Observatorio Astron\'omico Nacional,  M\'exico. An additional PN and one H\,{\sc ii} region were observed with this same telescope-spectrograph in 2013. Thus, in total we have high-quality data for 10 of the 26 PNe detected in this galaxy. In the wavelength calibrated spectra, the heliocentric radial velocities were measured with a precision better than 5-6 km s$^{-1}$. Data for two additional H\,{\sc ii} regions and some A-type supergiant stars were collected from  the literature. The heliocentric  radial velocities of the different objects were compared to the velocities of the H\,{\sc i} disk at the same position.}
   {From the analysis of radial velocities we found that H\,{\sc ii} regions and  A-type supergiants do share the kinematics of  the  H\,{\sc i} disk at the same position, as expected for  these young objects. In contrast,  most planetary nebula velocities differ significantly (more than 12 k m s$^{-1}$) from that of the H\,{\sc i} at the same position. The  kinematics of planetary nebulae is different from the young population kinematics and is more similar to the behavior shown by  carbon stars,  which are intermediate-age members of  the stellar spheroid existing in this galaxy. Our results confirm that there are at least two very different kinematical systems in NGC\,6822.
   }
   {}


   \maketitle
%

\section{Introduction}

   \label{sec:intro}
 Planetary nebulae (PNe) are produced in the advanced evolutionary stages of stars with initial masses from about 1 $M_{\odot}$ to  8 $M_{\odot}$ that have a wide age spread (from 0.1 to 9 Gyr, Allen et al. 1998). Therefore PNe are valuable tracers of  low and intermediate mass stars (LIMS). Because of their selective emission in a small number of strong and narrow emission lines, PNe can be discovered at significant distances within the nearby Universe (at least 30 Mpc). Studying them provides accurate information on the luminosity, age, metallicity, and dynamics of the
parent stellar population. This makes them very useful for testing several of theories about the evolution of low-intermediate mass stars and galaxies. In this work we study the kinematics of the PN population of the dwarf irregular galaxy NGC\,6822 to explore the connection of its intermediate-old stellar population with its H\,{\sc i} envelope.  A similar study on the relation of the PN population and the H\,{\sc i} gas envelope in other Local Group member, IC\,10, was carried out  by Gon\c{c}alves et al. (2012), who  reported that there is a kinematical connection between both  populations in IC\,10. Several authors have also used the PN population to analyze the kinematics of elliptical galaxies, for example,  Mc Neil-Moylan et al. (2012),  Teodorescu et al. (2011), and Coccato et al. (2009), who have analyzed PNe kinematics in ellipticals at about 20 Mpc, and  Ventimiglia, Arnaboldi \& Gerhard (2011) and Gerhard et al. (2007), who have studied ellipticals at 50-100  Mpc.

NGC\,6822 is a member of the Local Group, located at a distance of 459$\pm$10 kpc (Mateo 1998; Gieren et al. 2006). Its optical structure was discussed by Hodge (1977). It is dominated by a bar of about $8\arcmin$ long (equivalent to 1.07 kpc) and with a position angle (P.A.) of 10$^{\circ}$.  There is clear evidence of recent star formation in this galaxy, which includes more than a hundred  detected H\,{\sc ii} regions (Killen \& Dufour 1982; Hodge et al. 1988). A huge  H\,{\sc i} disk of about 6$\times$13 kpc, at P.A. of 110$^{\circ}$, centered on the  galactic center and including the optical bar,  was discovered recently (de Blok \& Walter 2000;  Weldrake et al. 2003).  In addition, by studying  carbon stars, Letarte et al. (2002) revealed that NGC\,6822 is surrounded by a huge stellar structure, composed mainly of intermediate-old age stars. de Blok \& Walter (2006), by analyzing the stellar component, also found that the old and intermediate age stars are significantly more extended than the H\,{\sc i} disk. The stellar spheroid has been mapped by Battinelli et al. (2006), who reported  a semi-major axis of 36$\arcmin$ (equivalent to 4.82 kpc) with a P.A. of 64.5$^{\circ}$; therefore the old stellar distribution is substantially different from the H\,{\sc  i} disk. With these characteristic, NGC\,6822 has been cataloged as a polar-ring galaxy (Demers et al. 2006) and it contains at least two dynamical systems: the huge H\,{\sc i} disk whose kinematics has been studied in depth by de Blok \& Walter (2000, 2006) and which shows a disk rotation, and the intermediate-age population, represented by carbon stars of the spheroid, whose kinematics was analyzed by Demers et al. (2006). This kinematics contrasts strongly with that of the  H\,{\sc  i} disk, because the spheroid rotates around its minor axis, whose P.A. differs by about 46$^{\circ}$ of the rotation axis of the H{\sc i} disk. 

 The main physical properties of NGC\,6822 are listed in Table 1.

\begin{table}[!th]
  \caption{Physical properties of NGC 6822} 
  \label{tab:prop}
 \begin{tabular}{lcc}
    \hline
   Property & Value & Ref.$^{a}$ \\
    \hline
   Hubble type & IB(s)m &  (1) \\
   Other ID  & DDO\,209, IC\,4895 &  \\
   R. A. (J2000)$^b$ & 19h 44m 56.6s & (2) \\
   Dec (J2000)$^b$ & -14$^{\circ}$ 48' 04.5'' & (2)  \\
   Distance (kpc) & 459$\pm$17 & (3) \\
   Systemic Vel. (km s$^{-1}$)  & -57$\pm$2, 55$\pm$2 & (4,5) \\
   Inclination (deg)  & 50.1 & (2) \\  
   E(B-V)  & 0.231 & (3) \\ 
   M$_{V}$  & -15.2$\pm$0.2 & (6) \\ 
   M$_{\star}$ ($10^{6}M_{\odot}$) & 100 & (4) \\ 
   M$_{HI}$ ($10^{6}M_{\odot}$) & 130 & (7)  \\
   12 + log(O/H), A & 8.36$\pm$0.19 & (8) \\
   12 + log(O/H), HII & 8.11$\pm$0.10 & (9) \\
  \hline
  \multicolumn{3}{l}{$^a$ {\small Ref.: (1) NASA/IPAC Extragalactic Database; (2) Brandenburg}} \\
\multicolumn{3}{l}{{\small   \& Skillman 1998; (3) Gieren et al. 2006; (4) Koribalski et al. 2004;   }}\\
\multicolumn{3}{l}{ {\small (5) Mateo 1998; (6) Dale et al. 2007; (7) Weldrake et al. 2003; } }\\
\multicolumn{3}{l}{{\small (8) Venn et al. 2001; (9) Hern\'andez-Mart{\'\i}nez et al. 2009}}\\
  \multicolumn{3}{l}{$^b$ {\small HI dynamical center coordinates.}}\\
    \end{tabular}
\end{table}

It has been proposed that dwarf irregular galaxies are dominated, at all radii, by dark matter that largely controls their kinematics. This seems to be the case of NGC\,6822 ( Weldrake et al. 2003; Hwang et al. 2014). 
In addition, numerical simulations based on the $\Lambda$CDM cosmology allow determining the dark-matter distribution and predict galactic dark-matter halos with cuspy density
profiles in dwarf irregular galaxies (Navarro et al. 1996). However, some galaxies seem to
favor the existence of  halos with soft cores (see the discussion in  Valenzuela et al. 2007 and references therein). Therefore these galaxies are very well-suited  laboratories for testing $\Lambda$CDM models. These phenomena can be traced through the analysis of the velocity
field of different objects in the galaxy, which provides some constraints that help to test the
models. Valenzuela et al. (2007) suggest that the kinematics of PNe, being intermediate age objects,
provides valuables clues to determine the shape of the dark matter
halo. The main aim in this work is to study the kinematical behavior of the
different types of populations existing in the dwarf irregular NGC\,6822  via the analysis 
of some  PNe,  H\,{\sc  ii} regions, and members of the  stellar components. In the long term, we will aim to discuss these kinematics in relation to $\Lambda$CDM models.

In \S \ref{sec:obs} we present the observations, data reduction, and data compiled from the literature, as well as the radial velocities derived for the different objects. In \S 3 the velocities of  PNe, H\,{\sc  ii} regions and two A-type supergiant stars are compared to the velocities of the H\,{\sc  i} disk. The results of the kinematical analysis are presented in \S 4. In \S 5 we  present the line profiles and the nebular diagnostic of some of our PNe as derived from our high resolution spectra. General results are presented in \S 6.

\section{Data acquisition and measurements of radial velocities}
\label{sec:obs}
\begin{table}[!t]
  \caption{PNe, HII regions, and stars analyzed in NGC 6822} 
  \label{tab:id_pn}
 \begin{tabular}{lrrc}
    \hline
    Object ID\tablefootmark{a} & \multicolumn{1}{c}{ RA (J2000) } & \multicolumn{1}{c}{Dec (2000)} & \multicolumn{1}{c}{Other ID }\\
    \hline
    PN2\tablefootmark{b} & 19:45:56.6 & -14:40:53.4 & PN5\tablefootmark{d}\\
    PN4\tablefootmark{b}\tablefootmark{c}& 19:45:01.4 & -14:41:33.0 & S30\tablefootmark{e} PN4\tablefootmark{d}\\
    PN6\tablefootmark{b}\tablefootmark{c}  & 19:44:00.7 & -14:42:43.6 & PN1\tablefootmark{d} \\
    PN7\tablefootmark{c} &19:44:49.1  &  -14:43:00.6  &PN2\tablefootmark{d} \\ 
    PN8\tablefootmark{b} &19:42:02.2 & -14:43:42.2 & PN6\tablefootmark{d} \\
    PN10\tablefootmark{c} & 19:45:22.0 & -14:43:28.0 & PN19\tablefootmark{d}\\
    PN12\tablefootmark{c} & 19:44:58.8 & -14:44:45.0 & S14\tablefootmark{e}, PN14\tablefootmark{d}\\
    PN14\tablefootmark{b} \tablefootmark{c}& 19:45:07.5 & -14:47:35.8 & S33\tablefootmark{e} ,PN7\tablefootmark{d}\\
    PN16\tablefootmark{c} & 19:44:58.6 & -14:46:10.0 & S16\tablefootmark{e}, PN13\tablefootmark{d}\\
   PN20\tablefootmark{b} & 19:45:11.5  & -14:48:53.6  & PN20\tablefootmark{f}\\
    \hline
    H\,V\tablefootmark{g} & 19:44:52.4 & -14:43:13.4 & --- \\
    H\,X\tablefootmark{g} & 19:45:5.23 & -14:43:16.7 & --- \\
     H\,{\sc  ii} 18\tablefootmark{c} & 19:44:42.6 & -14:50:30.2 & H{\sc ii}\,18\tablefootmark{d}, S10\tablefootmark{e}\\
     H\,{\sc  ii} 15\tablefootmark{c} & 19:44:57.2 & -14:47:50.6 & H{\sc ii}\,08\tablefootmark{d},  S28\tablefootmark{e}\\
     A\,13\tablefootmark{h}  & 19:44:53.4 & -14:46:42 &A supergiant\\
     A\,101\tablefootmark{h} & 19:44:56.5 & -14:46:14 & A supergiant \\
    \hline
     \end{tabular}
\tablefoottext{a}{Names as in Hern\'andez-Mart\'inez \& Pe\~na (2009), except for H\,V and H\,X.}\\
 \tablefoottext{b}{Observed at LCO with MIKE.} \\
  \tablefoottext{c}{Observed at SPM with MES.}\\
  \tablefoottext{d}{Leisy et al. (2005).}\\
  \tablefoottext{e}{Killen \& Dufour (1982).}\\
    \tablefoottext{f}{Richer \& McCall (2007).}\\
    \tablefoottext{g}{Hubble (1925).}\\
    \tablefoottext{h}{Venn et al. (2001).}
\end{table}

High spectral resolution data for four PNe (PN2, PN4, PN6, and PN14 of Table 2) were obtained  with the double echelle Magellan Inamori Kyocera Echelle spectrograph (MIKE) attached to the Clay Magellan Telescope at Las Campanas Observatory (LCO, Carnegie Institution, Chile), during the nights 2010 June 5 and 6. Two more objects (PN8 and PN20) were observed in 2014 March 6 and 8. The MIKE spectrograph operates with two arms that allow to obtain a blue and a red spectrum simultaneously (Berstein et al 2003). The standard grating setting was used, which provides a  full wavelength coverage   from  3350 \AA~ to 5050 \AA~ in the blue, and from 4950 \AA~ to 9400 \AA~ in the red.  The slit size was 1$''$ along the dispersion axis and 5$''$ in the spatial direction.  A binning of 2$\times$2 was used, thus obtaining a spacial scale of 0.2608 $''$/pix, while the spectral resolution  varied from 0.14 \AA~to 0.17 \AA~(about 10.8 km s$^{-1}$) in the blue and from 0.23 \AA~to 0.27 \AA~(about 12.8 km s$^{-1}$) in the red, as measured from the HWHM of the lines of the Th-Ar comparison lamp. Each object was observed on one occasion  with an exposure time of  15 minutes.  During the observing runs the seeing was better than 1$''$, most of the time. 

For the other three PNe,  high spectral resolution data  were retrieved from the database of the San Pedro M\'artir (SPM) Catalog of Extragalactic Planetary Nebulae (http://kincatpn.astrosen.unam.mx/, for details see Richer et al. 2010) where data for some objects in common with LCO observations were found as well. The data  available in this catalog consist of velocity-position diagrams for the [\ion{O}{iii}]$\lambda$5007 and H$\alpha$ emission lines, obtained with the Manchester Echelle Spectrometer (MES-SPM). In this case the spectral resolution  is 0.077 \AA/pix at [\ion{O}{iii}]$\lambda$5007 and 0.100\AA/pix at H$\alpha$, equivalent to about 11 km s$^{-1}$ for 2.6 pix FWHM. We downloaded the fits files for these PNe and calculated their heliocentric radial velocities from both lines, when available. An additional planetary  nebula (PN7) and the H{\sc ii} region \#15 were observed with the same instrument, MES, at SPM Observatory on July 4, 2013.  For the PN four spectra of 1800 s exposure time  each were acquired;  for the H{\sc ii} region two spectra of 1200 s and 900 s were obtained. The   H$\alpha$ filter was used in both cases. The spectra were combined and reduced according to the standard process for MES data (see Richer et al. 2010).  

\begin{figure*}[ht!]
  \begin{center}
\includegraphics[width=2\columnwidth]{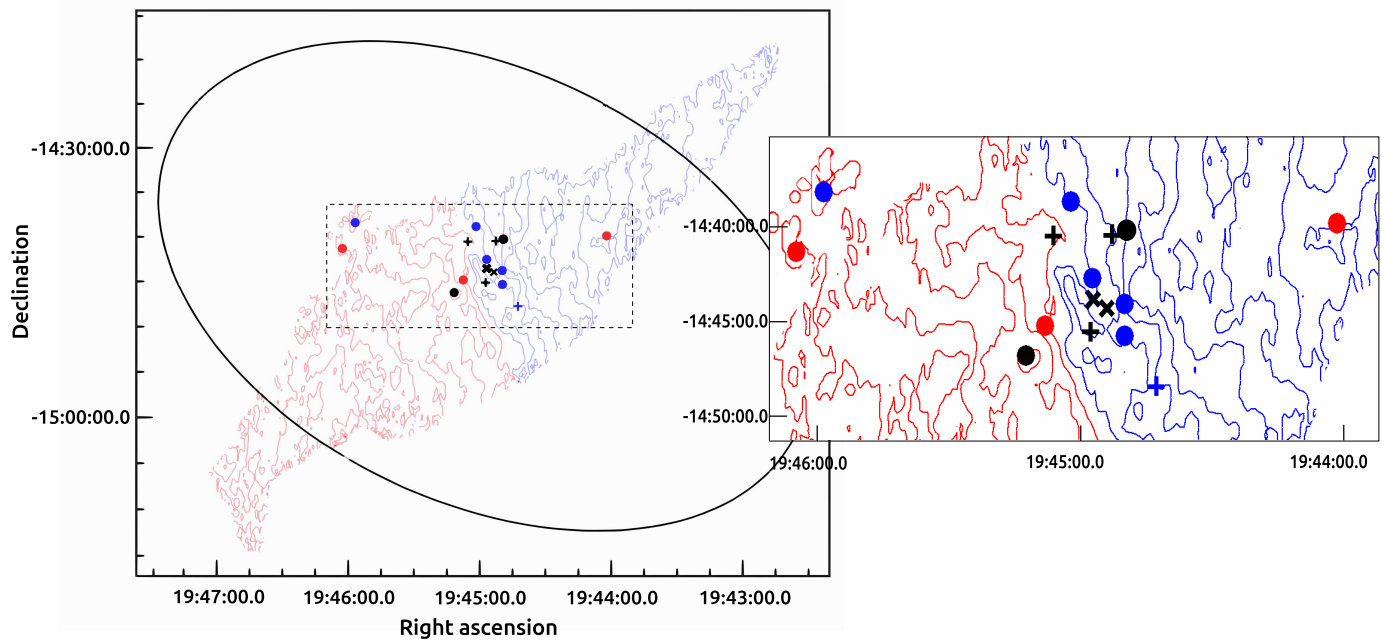}
  \caption{Radial velocity map of  H\,{\sc  i} from de Block \& Walter (2006).  The blue contours run from $-$55 km s$^{-1}$ in the inner part to $-$110 km s$^{-1}$ in the outer NW zone, in steps of 5 km s$^{-1}$. The red contours run from $-$50 km s$^{-1}$ in the central zone to +10 km s$^{-1}$ in the outer SE zone. The ellipsoid represents the stellar spheroid  analyzed by Battinelli et al. (2006) and Demers et al. (2006).  Position of the PNe (filled dots), H\,{\sc  ii} regions (crosses) and A-type supergiants (x marks) are those presented in Table~\ref{tab:id_pn}. Blue symbols indicate a negative difference  between observed objects and  H\,{\sc  i} velocities. Red symbols indicate a positive difference.  In black we plot objects with a velocity difference smaller than 12 km s$^{-1}$, relative to H\,{\sc i} disk.  A zoom of the central zone is shown.\label{fig:himap}}
\end{center}
\end{figure*}

Gathering all these data, we have information of the kinematical behavior of ten PNe, which represents almost 40 \% of the known PN candidates in this galaxy (26 objects reported by Hern\'andez-Mart{\'\i}nez \& Pe\~na 2009). It is important to note that we have data for several PNe near the galactic center, also for the PNe but that are the farthest away from the center (see Fig. ~\ref{fig:himap}).
\smallskip

In addition to PN data, spectroscopic data for some  H\,{\sc  ii} regions (which represent the youngest population in the galaxy) were  also obtained as another test to check the kinematical behavior of objetcs in NGC\,6822. High-resolution calibrated spectra for the bright H\,{\sc ii} regions H\,V and H\,X, obtained at the ESO Very Large Telescope with the UVES spectrograph at 2\AA~resolution, were kindly provided by A. Peimbert (Peimbert et al. 2005). Data for  H\,{\sc  ii}\,18  was extracted from the  SPM catalog mentioned above, and a new spectrum for  H\,{\sc  ii}\,15 was obtained as described in the previous paragraph.

On the other hand, high resolution spectra of two A-type supergiant stars (located near the center of the galaxy) were analyzed in depth by Venn et al. (2001). By performing spectral synthesis of several stellar lines, these authors   computed several stellar parameters, among them  accurate radial velocities that we used here (see their Table 1). These stars are considered as representative of the young stellar population.

 In next section (\S3) we compare the heliocentric velocities of PNe,  H\,{\sc  ii} regions and A-type supergiants  with the  H\,{\sc  i} gas velocities given by de Block \& Walter (2006) and with the kinematics of the carbon stars (Demers et al. 2006). 

In Table~\ref{tab:id_pn}  the identification of the observed objects is presented. Names for PNe and H\,{\sc ii} regions (except those of H\,V and H\,X) were adopted from Hern\'andez-Mart{\'\i}nez \& Pe\~na (2009). 

\subsection{Data reduction}

Data reduction of LCO Clay-MIKE spectra was carried out by using IRAF\footnote{IRAF is distributed by the National Optical Astronomy Observatories, which is operated by the Association of Universities for Research in Astronomy, Inc., under contract to the National Science Foundation.} echelle reduction packages. Raw 2D echellograms were bias-subtracted and flat-fielded. The  spectra were extracted with an extraction window of 3.72$''$ wide. Wavelength calibration was performed with a Th-Ar lamp, which was observed immediately after each science exposure. Data from 2010 run were flux calibrated using the spectro-photometric standard stars  Feige\,110, LDS\,749B, and NGC\,7293 (Oke 1990). No flux calibration was performed for the data obtained in March 2014. The full description of the reduction procedure can be found  in Garc{\'\i}a-Rojas et al. (2012). Heliocentric radial velocity correction was applied to all the spectra. 

For the MES-SPM data reduction procedure see Richer et al. (2010).  The fits files retrieved from  the SPM catalog provide wavelength calibrated spectra. These data are not flux calibrated. For these spectra the heliocentric velocity correction was also applied.  

\begin{table}[!th]\centering
  \caption{LCO PNe dereddened\tablefootmark{a} fluxes and heliocentric velocities for each line\tablefootmark{b}} 
  \label{tab:vpn}
 \begin{tabular}{cccccc}
    \hline
   Object  & \multicolumn{1}{c}{$\lambda_{obs}$ } & Ion & \multicolumn{1}{c}{$\lambda_{rest}$ }  & \multicolumn{1}{c}{Flux/ } & \multicolumn{1}{c}{$V_{helio}$ } \\
      & \multicolumn{1}{c}{(\AA)} &  & \multicolumn{1}{c}{(\AA)} &  \multicolumn{1}{c}{F(H$\beta$)}& \multicolumn{1}{c}{(km\: s$^{-1}$)}\\
    \hline
   PN2 & 4339.23 & $H \gamma$ & 4340.47 & 0.445 & -85.9\\
   PN2 & 4362.1: & [\ion{O}{iii}] & 4363.21 & 0.178 & -75.8:\\
   PN2 & 4859.96 & $H \beta$ & 4861.33 & 1.000 & -84.7 \\
   PN2 & 4957.55 & [\ion{O}{iii}] & 4958.91 & 1.395 & -82.4\\
   PN2 & 5005.47 & [\ion{O}{iii}] & 5006.84 & 4.142 & -82.3\\
   PN2 & 6560.93 & $H \alpha$ & 6562.82 & 2.812 & -86.4\\
   PN2 & 6581.59 & [\ion{N}{ii}] & 6583.41 & 0.072 & -83.0\\
   PN2 & 6716.1: & [\ion{S}{ii}] & 6716.47 & 0.032 & 19.1:\\
   PN2 & 6730.8: & \ion{[S}{ii]} & 6730.85 & 0.021 & -1.9:\\
   PN4 & 4339.47 & $H \gamma$ & 4340.47 & 0.500 & -68.9\\
   PN4 & 4362.19 & [\ion{O}{iii}] & 4363.21 & 0.224 & -70.1\\
   PN4 &  4684.70 & \ion{He}{ii} &  4685.68 & 0.229 &-63.4\\
   PN4 & 4860.25 & $H \beta$ & 4861.33 & 1.000 & -66.8\\
   PN4 & 4957.82 & [\ion{O}{iii}] & 4958.91 & 2.290 & -65.8\\
   PN4 & 5005.74 & [\ion{O}{iii}] & 5006.84 & 6.651 & -66.0\\
   PN4 & 6561.31 & $H \alpha$ & 6562.82 & 2.809 & -69.0\\
   PN4 & 6581.94 & [\ion{N}{ii}]  & 6583.41 & 0.394 & -66.9\\
   PN4 & 6715.23 & [\ion{S}{ii}] & 6716.47 & 0.029 & -55.3\\
   PN4 & 6729.41 & [\ion{S}{ii}] & 6730.85 & 0.059 & -64.0\\
   PN4 & 7134.06 & [\ion{Ar}{iii}] & 7135.78 & 0.091 & -72.4\\
   PN6 & 4339.37 & $H \gamma$ & 4340.47 & 0.468  & -76.3 \\
   PN6 & 4362.08 & [\ion{O}{iii}]& 4363.21 & 0.091  & -77.8\\
   PN6 & 4860.09 & $H \beta$ & 4861.33 & 1.000 &  -76.4\\
   PN6 & 4957.67 & [\ion{O}{iii}] & 4958.91 & 2.766 &  -75.1\\
   PN6 & 5005.61 & [\ion{O}{iii}] & 5006.84 & 7.257 &  -73.6\\
   PN6 & 6561.15 & $H \alpha$ & 6562.82 & 2.804 &  -76.6\\
   PN6 & 6581.78 & [\ion{N}{ii}] & 6583.41 & 0.412 & -74.3\\
   PN6 & 6714.66 & [\ion{S}{ii}] & 6716.47 & 0.017 & -80.9\\
   PN6 & 6729.05 & [\ion{S}{ii}] & 6730.85 & 0.235 & -80.3\\
   PN6 & 7133.94 & [\ion{Ar}{iii}] & 7135.78 & 0.525 & -77.0\\
   PN8 &  4685.29 & \ion{He}{ii} & 4685.68 &  --- &  -1.8\\
   PN8 & 4860.84 & $H\beta$ & 4861.33  & ---& -7.1\\
   PN8 & 4958.3: &  [\ion{O}{iii}]& 4958.91 & --- & -0.2: \\
   PN8  & 5006.35 & [\ion{O}{iii}] & 5006.84 & ---  & -6.2 \\
   PN8  & 6562.22 & $H\alpha$ & 6562.82 & --- & -4.3 \\
   PN14 & 4340.11 & $H \gamma$ & 4340.47 & 0.062 & -25.2\\
   PN14 & 4362.83 & [\ion{O}{iii}] & 4363.21 & 0.115 & -25.9 \\
   PN14 & 4685.33 & \ion{He}{ii} & 4685.68 & 0.463 &  -22.4 \\
   PN14 & 4860.88 & $H \beta$ & 4861.33 & 1.000 & -27.9\\
   PN14 & 4958.48 & [\ion{O}{iii}] & 4958.91 & 3.556 & -26.0\\
   PN14 & 5006.40 & [\ion{O}{iii}] & 5006.84 & 10.780 & -26.1\\
   PN14 & 6547.49 & [\ion{N}{ii}] & 6548.03 & 1.347 & -24.7\\
   PN14 & 6562.21 & $H \alpha$ & 6562.82 & 2.797 & -27.8\\
   PN14 & 6582.86 & [\ion{N}{ii}] & 6583.41 & 3.966 & -25.3\\
   PN14 & 6715.82 & [\ion{S}{ii}] & 6716.47 & 0.036  & -28.9 \\
   PN14 & 6730.25 & [\ion{S}{ii}] & 6730.85 & 0.039 & -26.7\\
   PN14 & 7135.13 & [\ion{Ar}{iii}] & 7135.78 & 0.064 & -27.5\\
   PN20 & 4860.40 & $H\beta$ & 4861.33 & --- & -33.8 \\ 
   PN20  & 4958.03 & [\ion{O}{iii}] & 4958.91 & --- & -29.7 \\
   PN20 & 5005.91 & [\ion{O}{iii}] & 5006.84 & --- & -32.2 \\
  \hline 
  \end{tabular}
\tablefoottext{a}{Fluxes dereddened by using Seaton (1979) reddening law.}
\tablefoottext{b}{A colon in the wavelength indicates a large uncertainty.}

\end{table}

\begin{table}[!t]\centering
  \small
  \newcommand{\DS}{\hspace{1\tabcolsep}} 
    \caption{Heliocentric velocities for SPM objects and comparison with LCO values} \label{tab:vhelio}
    \begin{tabular}{r @{\DS} cc l cc}
      \hline
      & \multicolumn{2}{c}{SPM}
      &&\multicolumn{2}{c}{LCO}\\
      & \multicolumn{2}{c}{(km s$^{-1}$)}
      &&\multicolumn{2}{c}{(km s$^{-1}$)}\\
      Object ID & \multicolumn{1}{c}{[\ion{O}{iii}]} & \multicolumn{1}{c}{H$\alpha$} && \multicolumn{1}{c}{[\ion{O}{iii}]} & \multicolumn{1}{c}{H$\alpha$} \\
             & \multicolumn{1}{c}{$\lambda$5007} & \multicolumn{1}{c}{$\lambda$6563} && \multicolumn{1}{c}{$\lambda$5007} & \multicolumn{1}{c}{$\lambda$6563} \\   
      \hline
      PN4  & $-59.9$ & $-64.3$ &&$-66.0$ & $-69.0$ \\
      PN6  & $-72.6$ & --- && $-73.6$ & $-76.6$ \\ 
      PN7 & ---  &  $-55.4$  & &--- & ---\\
      PN10 & $-69.6$ & $-70.9$ &&--- & --- \\
      PN12 & $-86.8$ & $-87.0$&& --- & --- \\
      PN14  & $-29.3$ & $-29.0$ && $-26.1$ & $-27.8$ \\
      PN16 & $-77.1$ & $-76.1$&& --- & --- \\
       H\,{\sc  ii}\,18 &$-88.1$ &---  && --- & --- \\
      H\,{\sc ii}\, 15 & --- & -73.3 & &--- & --- \\
      \hline
    \end{tabular}
\end{table}
\subsection{Measuring the radial velocities}
For all the objects observed at LCO, we systematically chose the most intense lines in the  blue and red spectra to measure velocities. They were measured by fitting a Gaussian profile to the observed  profile  using the {\it splot}  IRAF routine. In Table~\ref{tab:vpn} we present the lines measured for each object: in column 1 object  names are listed, the observed wavelength, $\lambda_{obs}$ are presented in column 2;  the ion identification is found in column 3, and in column 4 the laboratory wavelength is given;  the observed flux relative to H$\beta$ and the heliocentric radial velocity calculated are listed in columns 5 and 6.  We adopted the average velocity given by all the measured lines as the heliocentric velocity of the object. 

 For the objects from the SPM catalog, the available lines (H$\alpha$, [\ion{O}{iii}] 5007) were measured and the heliocentric radial velocity calculated. In Table~\ref{tab:vhelio} we present these  velocities  and compare them with the values from LCO observations for the objects in common. It can be seen that both sets of  values  are very similar.  The largest differences is found for PN4, where $\Delta$V between LCO and SPM data for [\ion{O}{iii}] $\lambda$5007 and H$\alpha$ lines are 6.1 km s$^{-1}$ and 4.7 km s$^{-1}$ respectively; these differences lie within the uncertainties. In all  the other cases the differences are smaller. This shows that our velocities obtained from the high resolution spectra, from LCO and SPM are very reliable because they have uncertainties smaller than about  5-6 km s$^{-1}$.
 
 For the H\,{\sc ii} regions H\,V and H\,X,  the central wavelengths of H$\beta$, and [\ion{O}{iii}]$\lambda\lambda$4959  and 5007 were measured by adjusting a Gaussian profile to the lines, to determine their  heliocentric radial velocities. The dispersion given by the lines is smaller than 3 km s$^{-1}$.

\begin{table*}[!th]\centering
  \caption{Heliocentric velocities of PNe, H\,{\sc ii} regions, A supergiants, and  H\,{\sc  i} gas} 
  \label{tab:vpn HII-HI}
 \begin{tabular}{lrrrrr}
    \hline
   Object  & \multicolumn{1}{c}{V$_{helio}$ } & \multicolumn{1}{c}{$\pm$ } & \multicolumn{1}{c}{V$_{H I}$ } &  $\Delta$~~~& $\Delta_{sys}$ \\
       & \multicolumn{1}{c}{(km/s)} & \multicolumn{1}{c}{(km/s)} & \multicolumn{1}{c}{(km/s)} & (km/s) & \multicolumn{1}{c}{(km/s)}  \\
    \hline
   PN2$^a$ & -84.1 & 1.7 & -24.0 & -60.1 & -29.1\\
   PN4   & -66.5 & 3.8 & -54.4 & -12.2 & -11.5\\
   PN6   & -76.8 & 2.1 & -89.2 &  +12.4 & -21.8\\
   PN7   & -55.4 & 4.6 & -64.9 & +9.5 & -0.4\\
   PN8   & -4.7 & 3.0 & -20.8 &  -16.1 & +50.2 \\
   PN10  & -69.6 & 3.2 & -54.1 &  -15.5 & -14.6\\
   PN12  & -86.7 & 4.9 & -65.8 &  -20.9 & -31.7\\
   PN14  & -26.5 & 1.3 & -47.5 &  +21.0 & +28.5 \\
   PN16  & -77.1 & 4.7 & -60.6 &  -16.5 & -22.1\\
   PN20  & -31.8 & 6.0 & -36.9 & +5.1 & +23.1 \\
   H\,V  & -62.7 & 2.1 & -58.2 &  -4.5  & -7.7\\  
   H\,X  & -57.3 & 2.5 & -48.7 &  -8.6 & -2.3\\ 
   H\,{\sc  ii} 18 & -87.4 & 1.8 & -63.2 &  -24.2 & -32.4\\ 
   H\,{\sc  ii} 15 & -73.0 & 6.0 & -61.3 &  -11.7 & -18.0\\ 
   A\,13$^c$ & -55 & 2.0 & -57.3 & 2.3 & 0.0\\
   A\,101$^c$ & -65 & 2.0 & -53.8 & -11.2& -10.0\\
  \hline
 \multicolumn{6}{l}{$^a$ The weak [\ion{S}{ii}]  and [\ion{O}{iii}]4363 lines were not considered.}\\
  \multicolumn{5}{l}{$^b$ Large error due to low resolution spectrum.}\\
  \multicolumn{5}{l}{$^c$ A supergiants analyzed by Venn et al. 2001.}
    \end{tabular}
\end{table*}

\section{Radial velocities of different objects}
In this section  the behavior of the radial velocities of the  objects is analyzed in relation to the velocities of the H\,{\sc i} disk and the stellar spheroid at the same projected position .
\subsection{H\,{\sc  i} disk}
Using the velocity field  obtained by de Block \& Walter (2006) for the  H\,{\sc  i} disk, we compared the  heliocentric velocities of  PNe, H\,{\sc ii} regions, and A-type supergiants  with those of the  H\,{\sc  i} gas. The results  can be found in Table~\ref{tab:vpn HII-HI} where we present for all  our objects the adopted heliocentric velocity and its uncertainty (at 1$\sigma$), and the velocity of the  H\,{\sc  i}  at the same projected position. The fifth column shows the difference $\Delta$ = V$_{helio}~-$ V$_{HI}$, and in the sixth column we present the difference between the  systemic velocity ($-$55 km s$^{-1}$ from Mateo 1998) and our objects. 

The position of our objects in the galaxy  and their difference in velocity relative to the HI disk are illustrated in Figure~\ref{fig:himap}, where we have included the location of the spheroid of intermediate-age  stars  (Battinelli et al. 2006; Demers et al. 2006), which has a semi-major axis of 36$\arcmin$ long, oriented at  P.A. of 64.5$^{\circ}$. \\

Column 5 of Table~\ref{tab:vpn HII-HI} show that the differences in velocity between the PNe (all except two) and the H\,{\sc i} at the same position, are larger than $\pm$12 km s$^{-1}$, which is significantly higher than the uncertainties. According to the velocity map for H\,{\sc i} by de Blok \& Walter (2006) (see Fig. 1), the  H\,{\sc  i} disk rotates around the optical center with its  NE zone  moving toward us (heliocentric velocities from $-$110 to $-$55 km$^{-1}$), while the SW zone  recedes from us at velocities from $-$50 to +10 km s$^{-1}$.  Half of our PNe are approaching faster than the  H\,{\sc  i} gas, with differences from  $-$60  km s$^{-1}$ to $-$12  km s$^{-1}$, while the other half are receding from the disk with velocity differences  from +5  km s$^{-1}$ to +21  km s$^{-1}$, that is, the H\,{\sc i} disk is leaving these objects behind. 

\begin{figure}[!th]
  \begin{center}
\includegraphics[width=1.0\columnwidth,height=7.5cm]{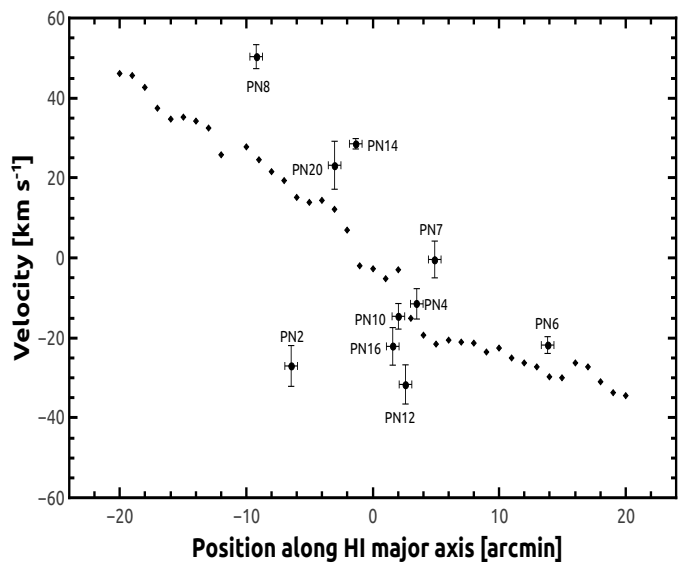}
  \caption{PNe velocities, relative to the system, and their positions  projected on the H\,{\sc i} disk major axis. The dotted line shows the rotation of the disk.  \label{fig:hipn}}
\end{center}
\end{figure}

 To better illustrate the differences in velocity between our PNe and the H\,{\sc i} disk, Fig. ~\ref{fig:hipn} shows the PN velocities (relative to the system) vs. their position projected along the major axis of the H\,{\sc i} disk. The rotation of the H\,{\sc i} disk is shown as a dotted line, and it shows some irregularities that so far have not been explained. However, this figure shows  that  PNe do not rotate with the disk, but have a much more disperse velocity field.

In contrast, the heliocentric velocities  of the H\,{\sc  ii} regions H\,V, H\,X and  H\,{\sc  ii}\,15  and the A-type supergiants are very similar to those of the  H\,{\sc  i} disk, with differences smaller  than 12  km s$^{-1}$,  as expected because they are young objects. They share the movement of the H\,{\sc i} disk.
In the case of  H\,{\sc  ii}\,18, which is cataloged as a  H\,{\sc  ii} region by Leisy et al. (2005), it presents a large velocity difference  of $-$24.2 km s$^{-1}$ relative to  the  H\,{\sc  i} gas. In the appendix we discuss the possibility that this nebula is in fact a planetary nebula superposed on a H\,{\sc  ii} region, which would explain its strange kinematics.

\subsection{Stellar spheroid and its velocity field}

\begin{figure}[!th]
  \begin{center}
\includegraphics[width=1.0\columnwidth,height=7.5cm]{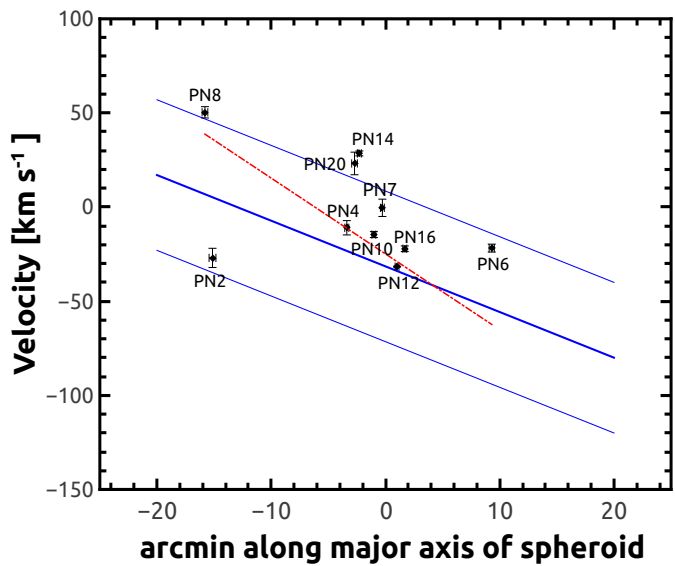}
  \caption{Systemic velocity of PNe vs. their position projected along the major axis of the spheroid. The central line is the best fit to the velocities of the carbon stars (relative to the systemic velocity), as given  by Demers et al. (2006). Top and bottom lines indicate their dispersion.  A linear fit to the PNe velocities is also shown (dashed-dotted red line). The velocity uncertainties of PNe are marked.  \label{fig:pndsh}}
\end{center}
\end{figure}

Demers et al. (2006) measured radial velocities of 110 carbon stars located within 15$\arcmin$ of the  H\,{\sc  i} major axis with a precision of $\pm$15 km s$^{-1}$. These stars belong to the stellar spheroid and seem to be rotating with a rotation axis that almost coincides with the minor axis of the stellar  spheroid. Thus, the rotation axis of the C-star system and that of the H\,{\sc i} disk differ by about 46$^{\circ}$, which indicates that there are two different kinematical systems.

 In the following we compare the PN heliocentric velocities with those of the carbon stars. We projected the positions of the PNe with respect to the major axis of the ellipsoid (P.A.~65$^{\circ}$) and, subtracting the system velocity, we located them in Figure~\ref{fig:pndsh}, where we show a band representing the average  behavior  of  carbon star velocities (and its dispersion). The uncertainties of our data in this figure are not larger than 6 km s$^{-1}$. {\bf We included a linear fit to the velocities of our PN sample.} 

Interestingly, we found that the heliocentric velocities of PNe are similar to those reported  for the carbon stars near the same position. Our PN sample also shows a velocity dispersion as large as the C stars. PN2 and PN8, located in the East part of the bar, show the strongest differences with the average of carbon star velocities, although it is always inside the C-star velocity dispersion.  The fit for the PNe shown in Fig.~\ref{fig:pndsh} seems steeper than that for the C-stars, but as this is a fit for only ten objects, it is not very reliable. However, we consider that both behaviors are similar in the sense that the PN system seems to rotate in a way similar  to the C-star spheroid. Nevertheless, one important difference is that the PN average velocity is -57.8 km s$^{-1}$, which agrees very well with the systemic velocity, while the C-star mean velocity is -32.9 km s$^{-1}$, that is, the  spheroid would be receding relative to the galaxy, with a velocity of about 22 km s$^{-1}$. This latter possibility was indicated by Hwang et al. (2014), who in addition, by studying the kinematics of four extended stellar clusters, found that these objects (located from one side to the other of the extended stellar spheroid) do not rotate as the C-stars, and they present an average velocity of  -88.3 km s$^{-1}$, which also very different from the system velocity (the stellar clusters would be moving towards us 30 km s$^{-1}$ faster than the galaxy). In both cases (C-stars and stellar clusters) this is a peculiar kinematical behavior that deserves future studies.



\section{Conclusions for the kinematics}
\label{sec:cons}

By using high spectral resolution data obtained at LCO and SPM, we studied the kinematics of ten PNe and four  H\,{\sc  ii} regions and compared them with the kinematics of the extended  H\,{\sc  i} disk  and the C stars of the stellar spheroid of NGC\,6822. The behavior of two  A-type supergiant stars was analyzed as well.


 It is clear that the studied  PNe are not moving along with the  H\,{\sc  i} gas and their kinematics is more similar to that of the C-stars of the stellar spheroid. This is not the case of the A-type supergiants and the H\,{\sc  ii} regions H\,V, H\,X, and  H\,{\sc  ii}\,15, which seems to be part of the dynamical system of the H\,{\sc i} disk, which is expected because they are young objects. 
 
 One of the regions that was previously declared a H\,{\sc  ii} region, the compact  H\,{\sc  ii}\,18, shows kinematics closer to the PNe and C stars, with a difference in velocity of $-$24.2 km s$^{-1}$ relative to the  H\,{\sc  i} gas at the same position. In the appendix we propose  that this nebula might be a true PN located near a faint and extended  H\,{\sc  ii} region, which might explain the fuzzy appearance of  H\,{\sc  ii}\,18. And this would explain  its peculiar kinematics.
 
Because of their systemic velocities,  PNe seem to be part of the intermediate-age stellar spheroid in NGC\,6822. However, the average velocity of our sample of ten PNe is -57.8 km s$^{-1}$, therefore the PN sample shares the system velocity of -55 km s$^{-1}$. On the other hand, the C-stars measured by Demers et al. (2006) show an average velocity of  -32.9 km s$^{-1}$ and the four stellar clusters reported by Hwang et al (2014) have an average velocity of -88.3 km s$^{-1}$.  Therefore, it seems  that NGC\,6822 has different kinematical systems (the H{\sc i} gas and the young objects, the stellar spheroid with its C-stars, the stellar clusters in the stellar halo, etc.) that deserve to be investigated in more detail. 

 \section{ Line profiles and other characteristics of analyzed PNe}

With the high spectral resolution used, the line profiles of the objects observed at LCO are resolved. In Figure 3 we present  the profiles of [\ion{O}{iii}] $\lambda$5007, H$\alpha$, and [\ion{N}{ii}] $\lambda$6583 lines for PN6, PN4, PN2 and PN14.
The profiles are different in each case: PN6 presents more compact lines, very Gaussian in shape. The FWHM of lines (after subtracting the instrumental and thermal widths by assuming that they add in quadrature) is 17$\pm$2 km s$^{-1}$, indicating an expansion velocity of around 8 km s$^{-1}$. The situation is different in PN4 where the lines are much wider and show a structure that seems to contain three components; this is  particularly clear in the [\ion{N}{ii}] line, where a central component surrounded by some kind of shell is apparent. The  shell would have an expansion velocity of about 25 km s$^{-1}$.
In PN2 the lines are Gaussian but wider than for PN6, with a FWHM of about 25 km s$^{-1}$. Then, the expansion velocity in PN2 would be of about 12 km s$^{-1}$.

PN 14 shows the most interesting profiles. The [\ion{O}{iii}] lines present two very close components differing in velocity by about 15 km s$^{-1}$, 
and in the [\ion{N}{ii}]$\lambda$6583 profile a faint very wide component is apparent. This latter component represents some bipolar ejection at high velocities (about 140 km s$^{-1}$) similar as the galactic M1-25 and M1-32 (among others, Medina et al. 2006; Akras \& L\'opez 2012).

\begin{figure*}[!th]
  \newlength\thisfigwidth
  \setlength\thisfigwidth{0.2\linewidth}
  \addtolength\thisfigwidth{-0.5cm}
  \parbox[t]{\linewidth}{%
     \vspace{0pt}
      \includegraphics[width=5.2cm,height=4.8cm]{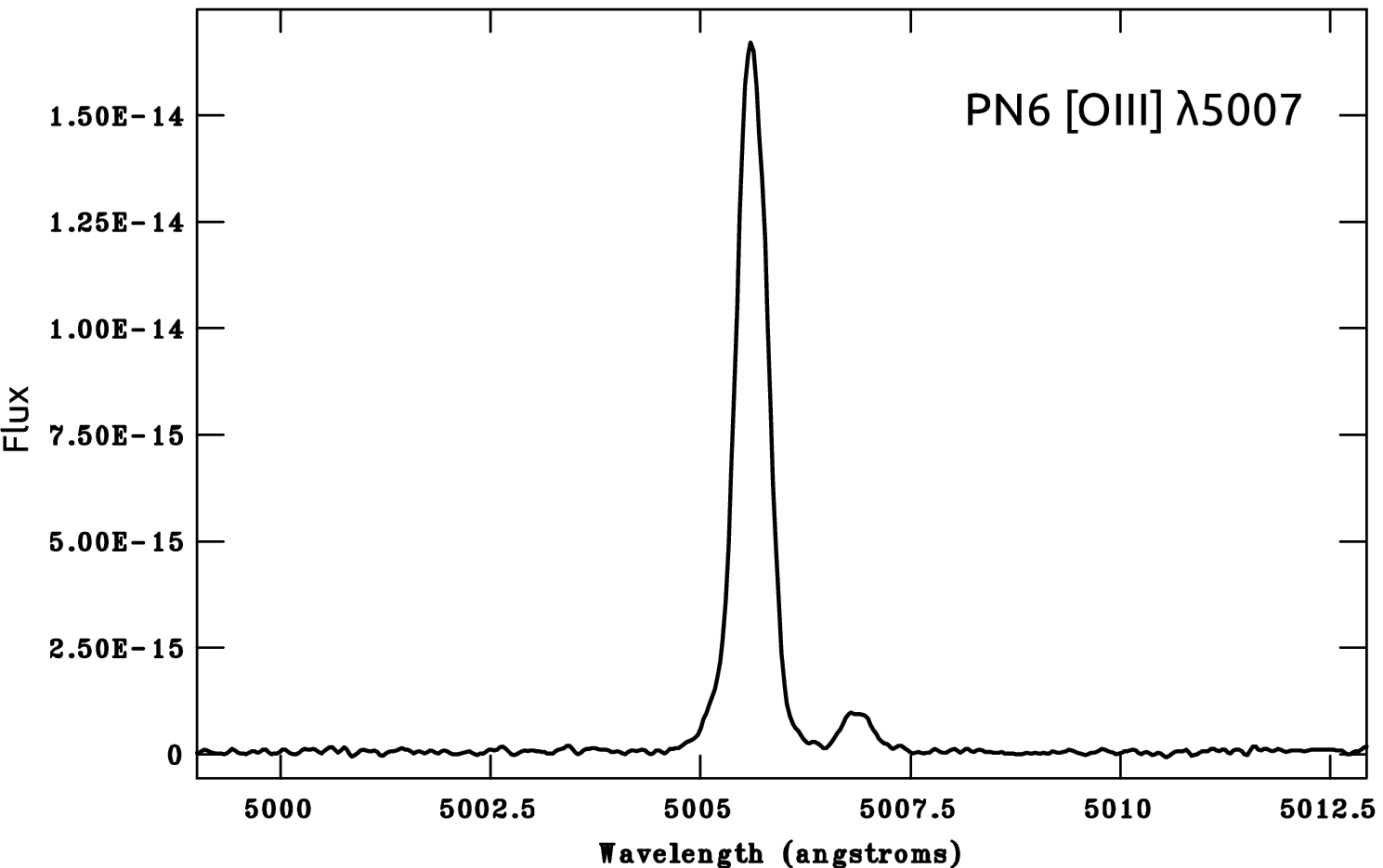}
     \hfill%
     \includegraphics[width=5.2cm,height=4.8cm]{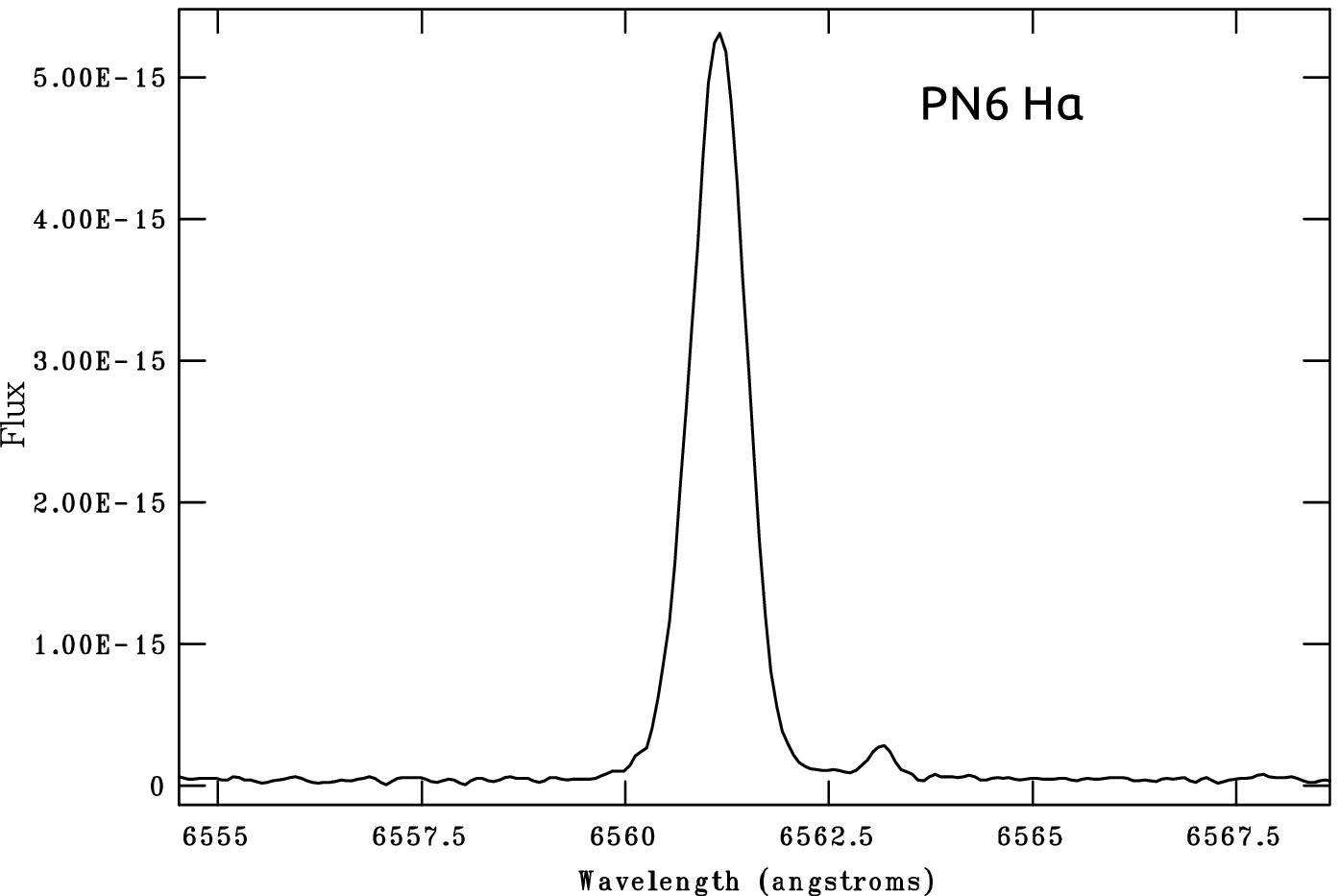}
     \hfill%
     \includegraphics[width=5.2cm,height=4.5cm]{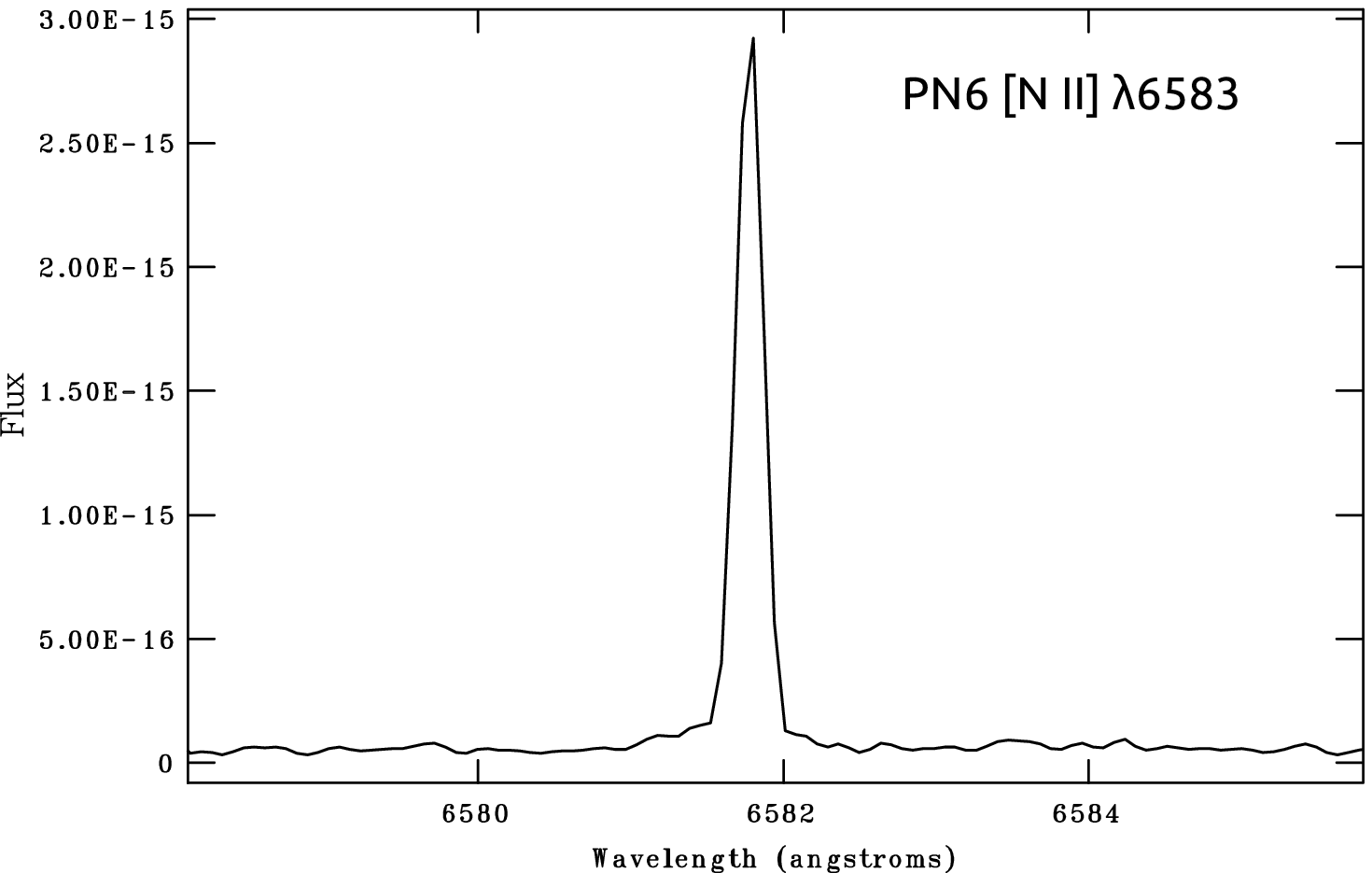}
     }
  \parbox[t]{\linewidth}{%
     \vspace{0pt}
     \includegraphics[width=5.2cm,height=4.4cm]{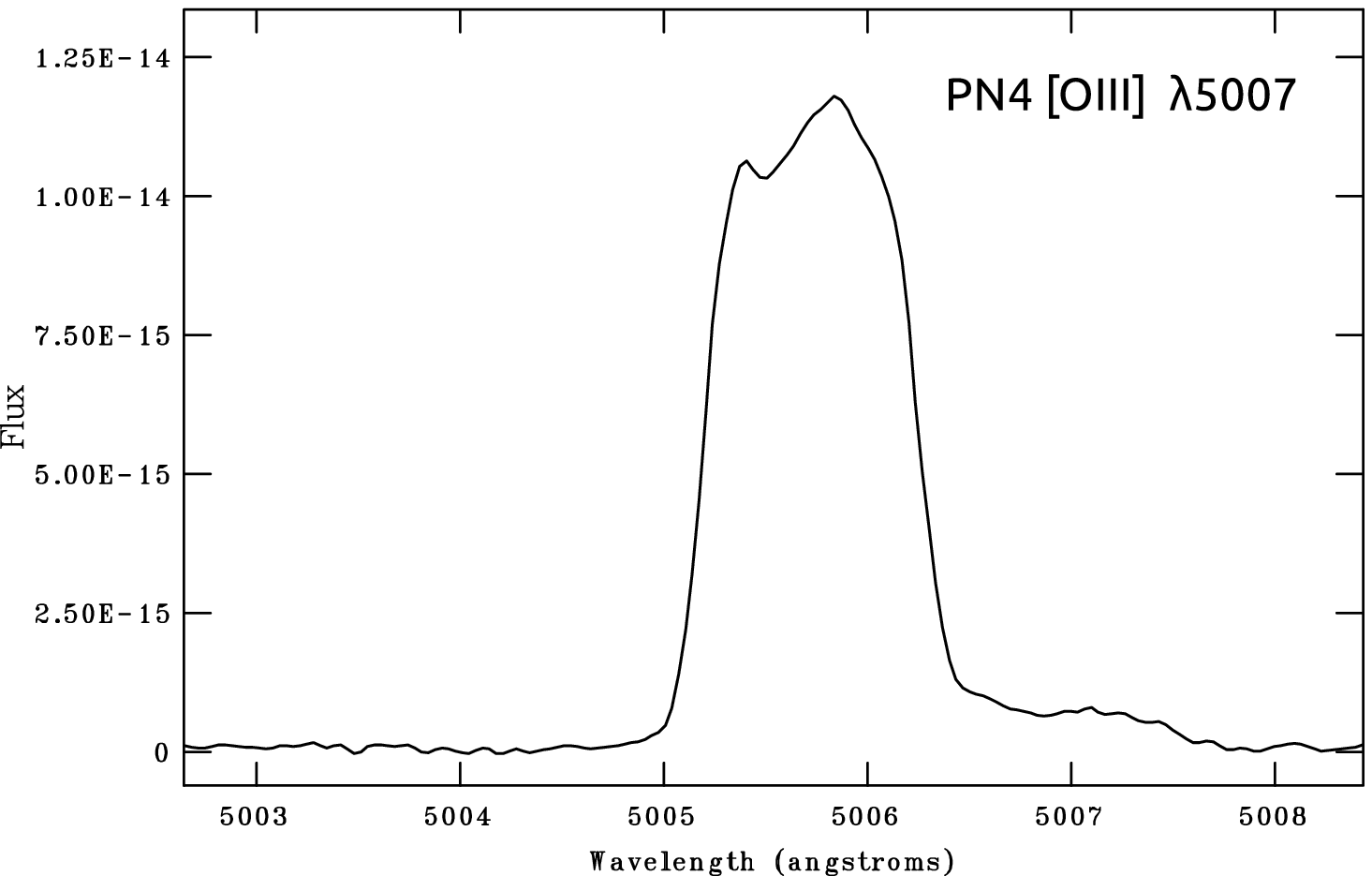}
     \hfill%
     \includegraphics[width=5.2cm,height=4.8cm]{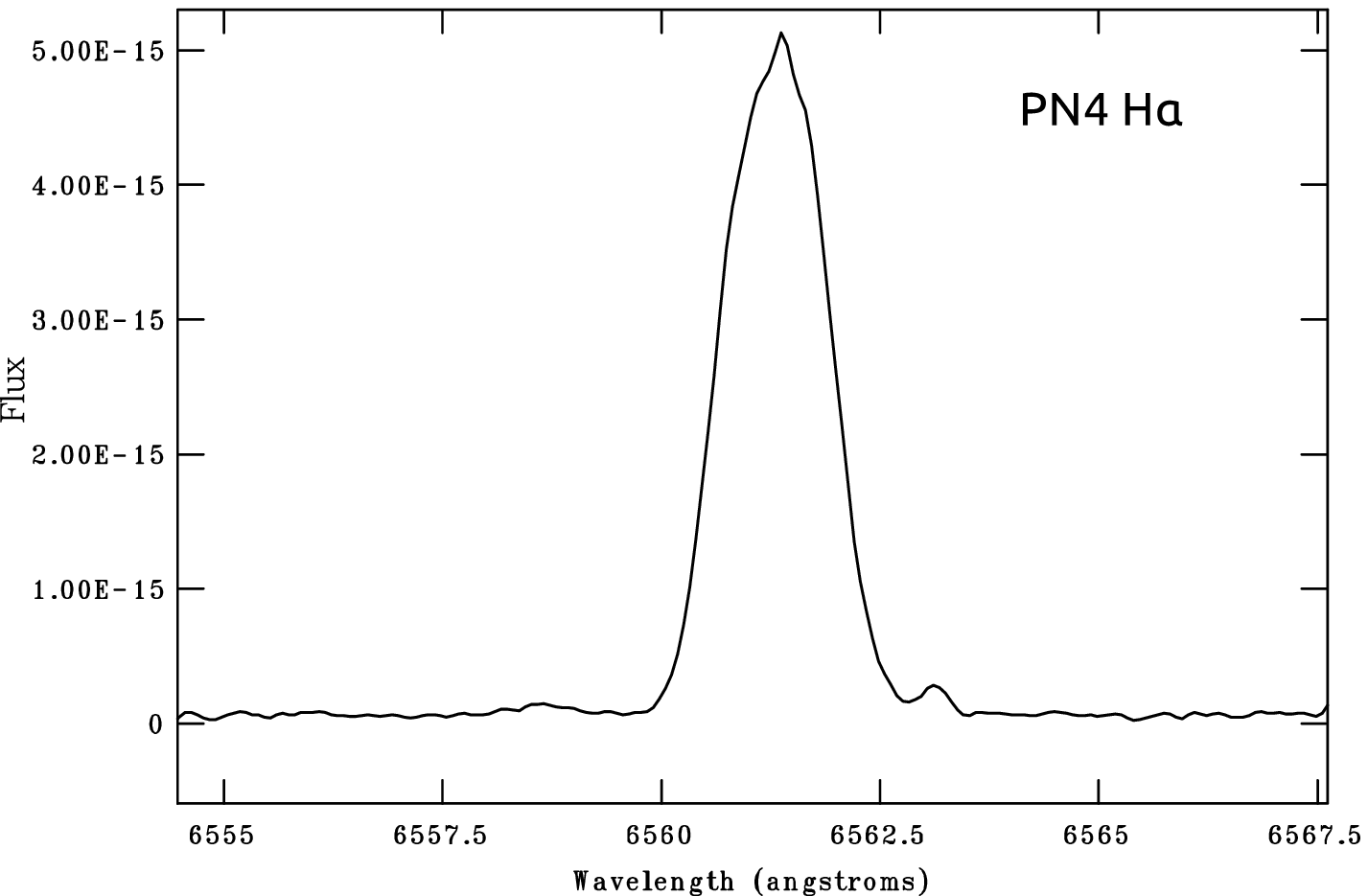}
     \hfill%
	  \includegraphics[width=5.2cm,height=4.8cm]{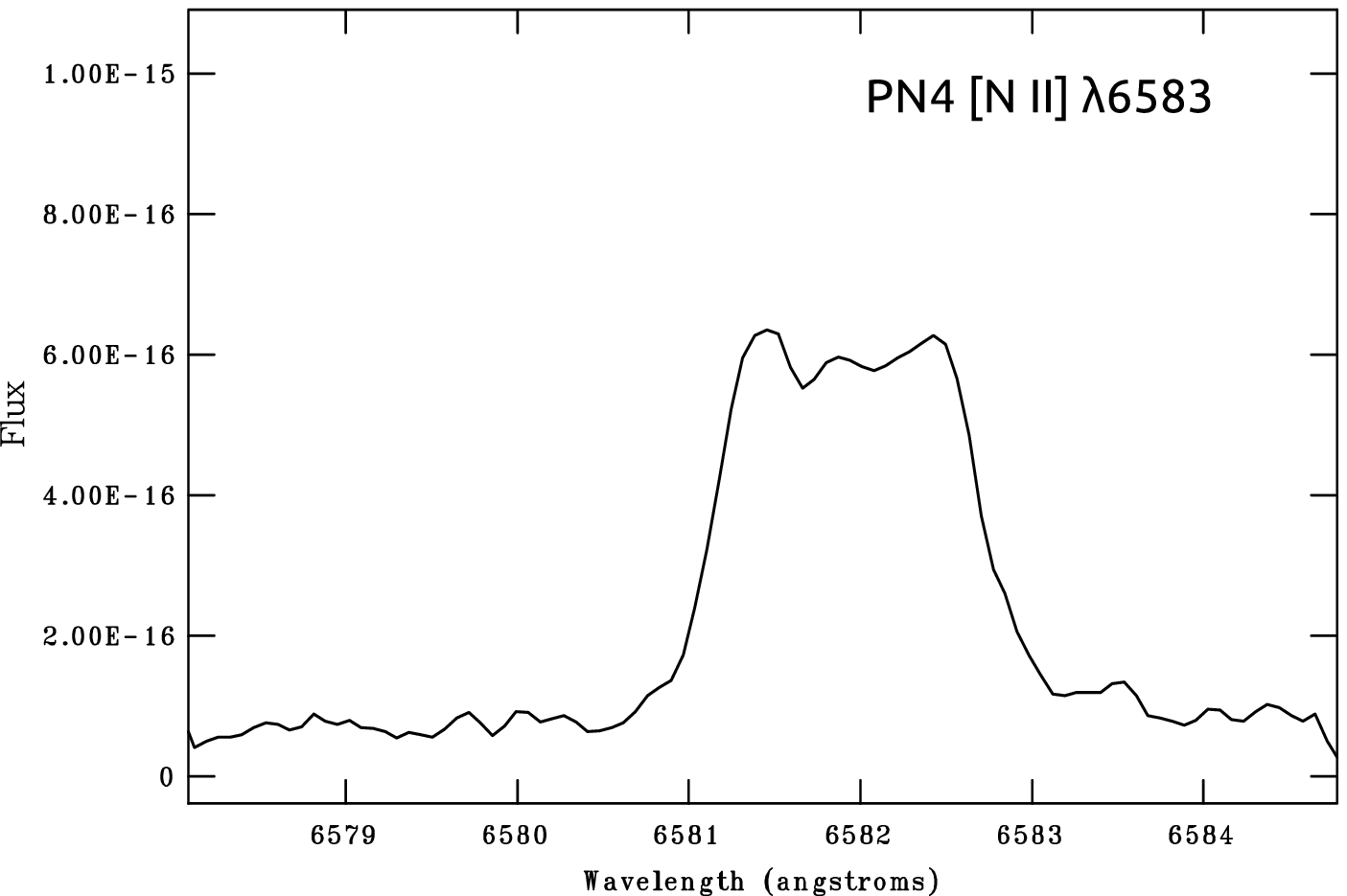}
     }
  \parbox[t]{\linewidth}{%
     \vspace{0pt}
	  \includegraphics[width=5.2cm,height=4.8cm]{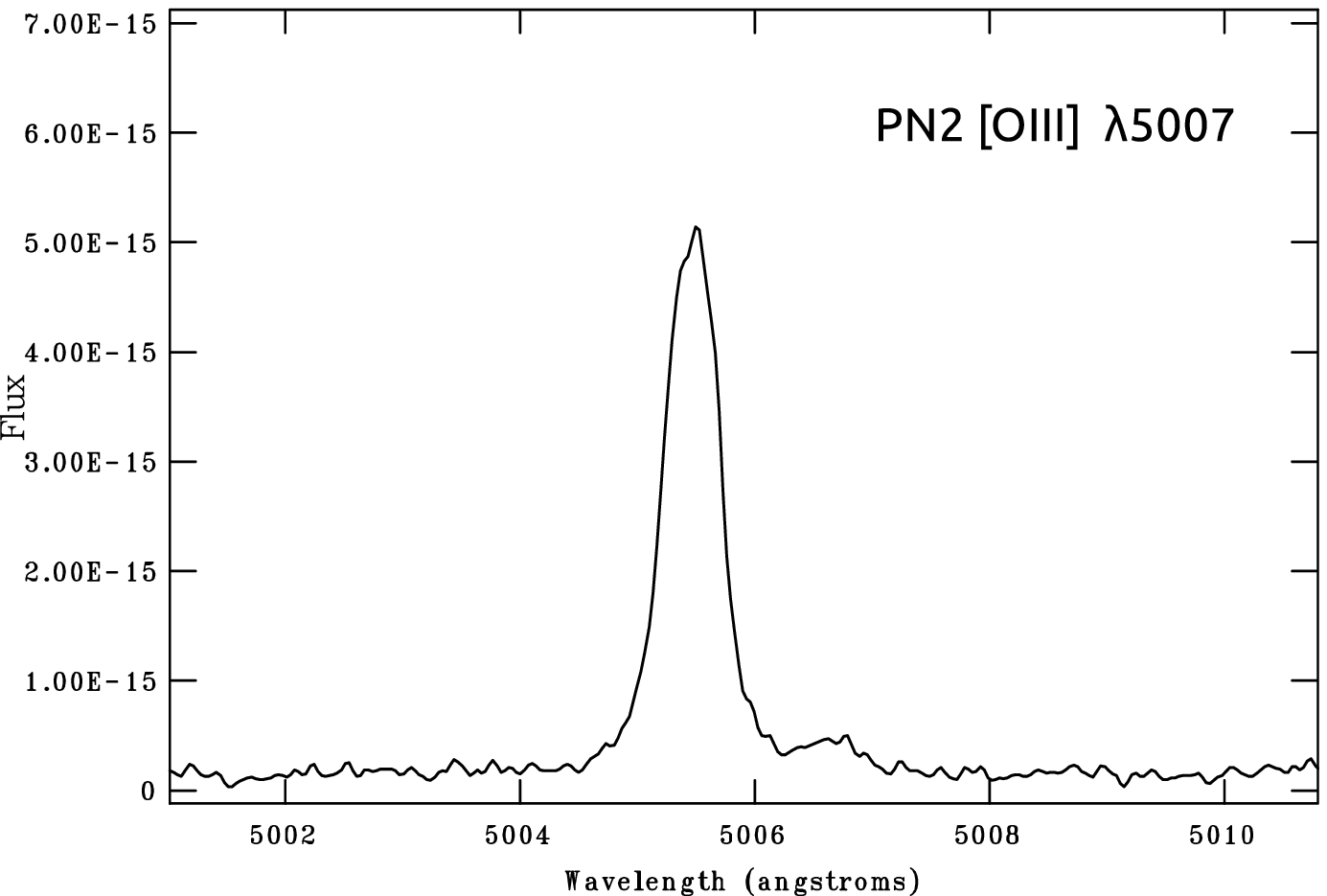}
     \hfill%
     \includegraphics[width=5.1cm,height=4.45cm]{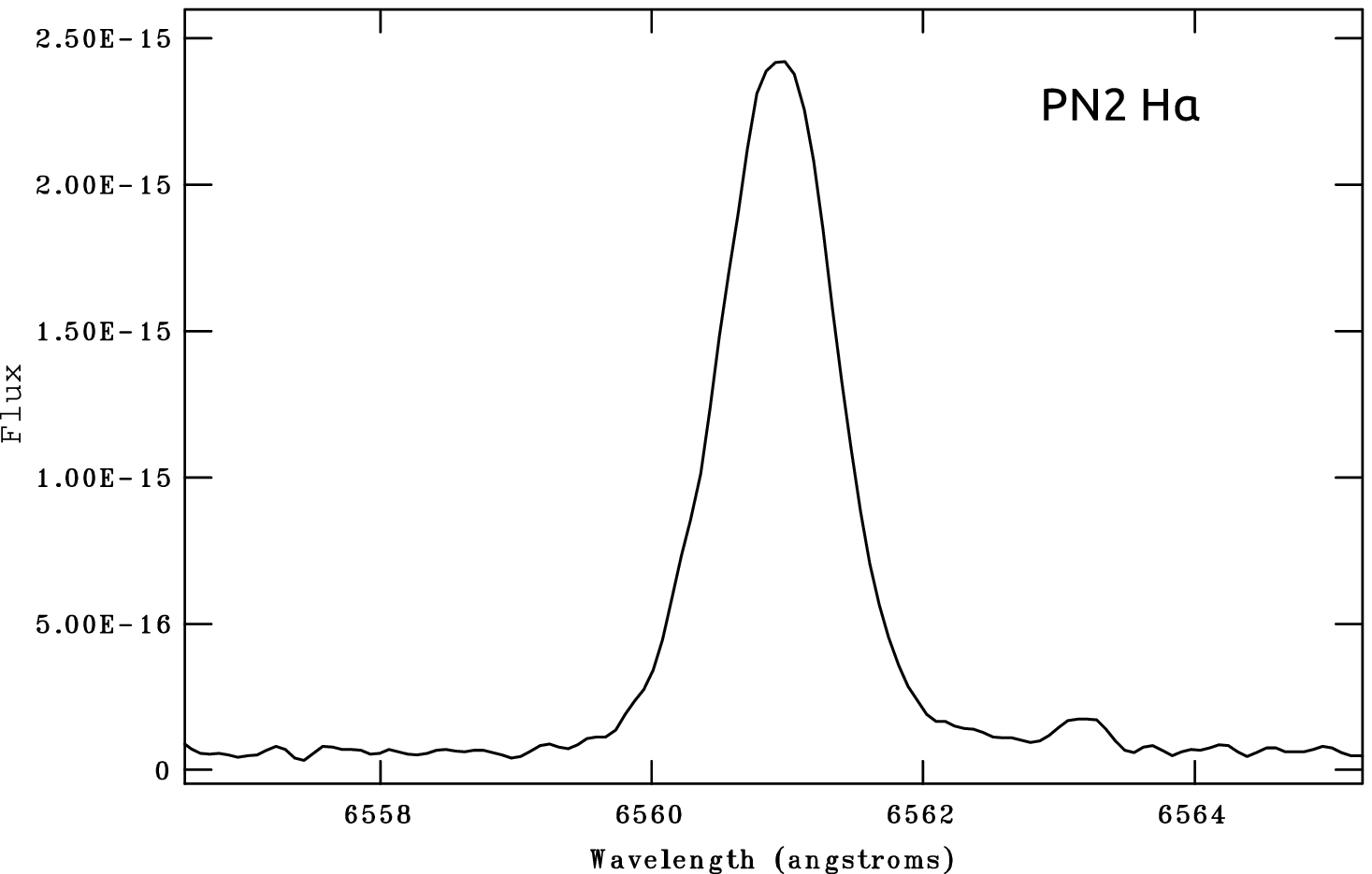}
     \hfill%
	  \includegraphics[width=5.2cm,height=4.5cm]{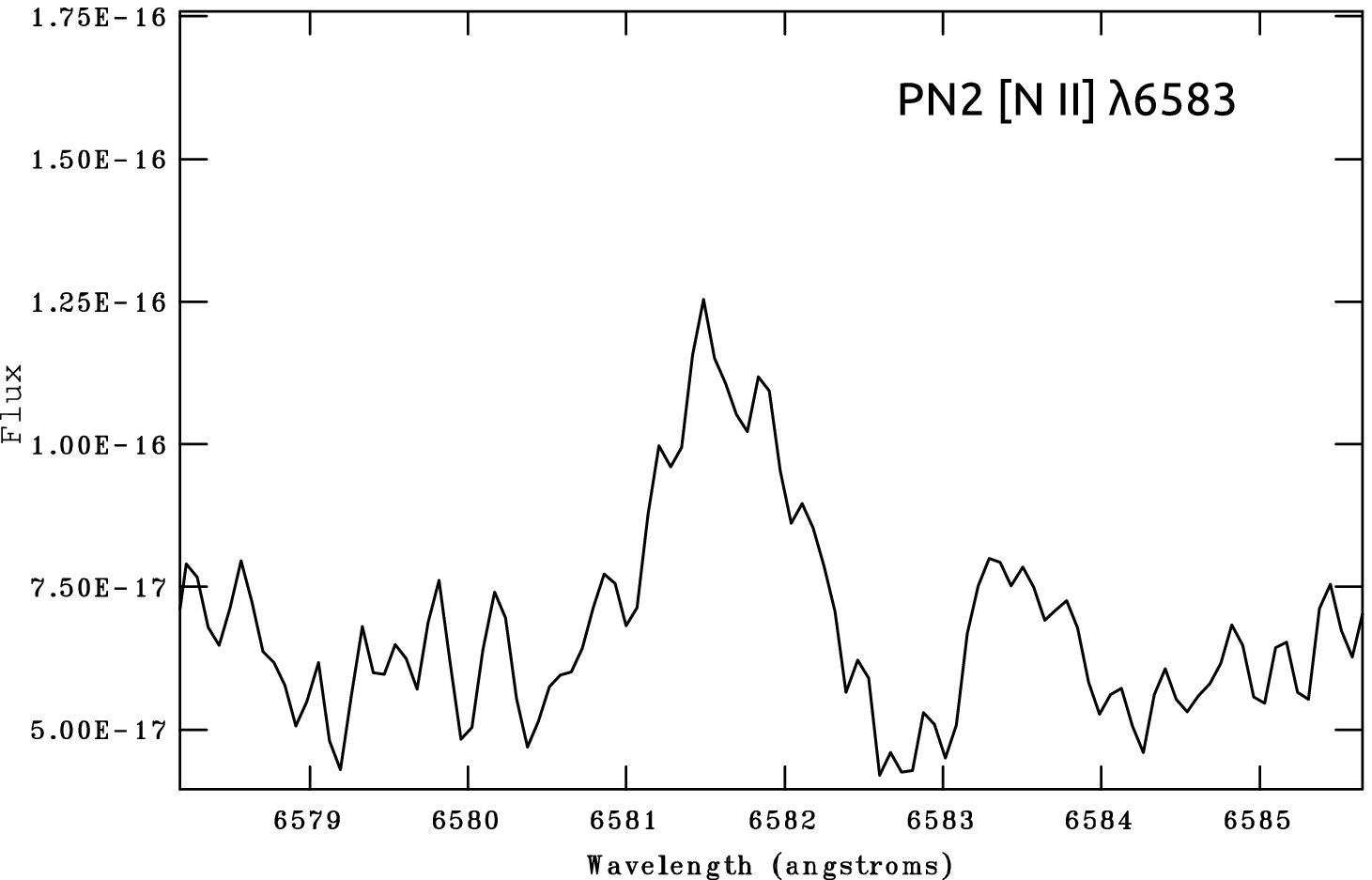}
 	 }
  \parbox[t]{\linewidth}{%
     \vspace{0pt}
	 \includegraphics[width=5.2cm,height=4.8cm]{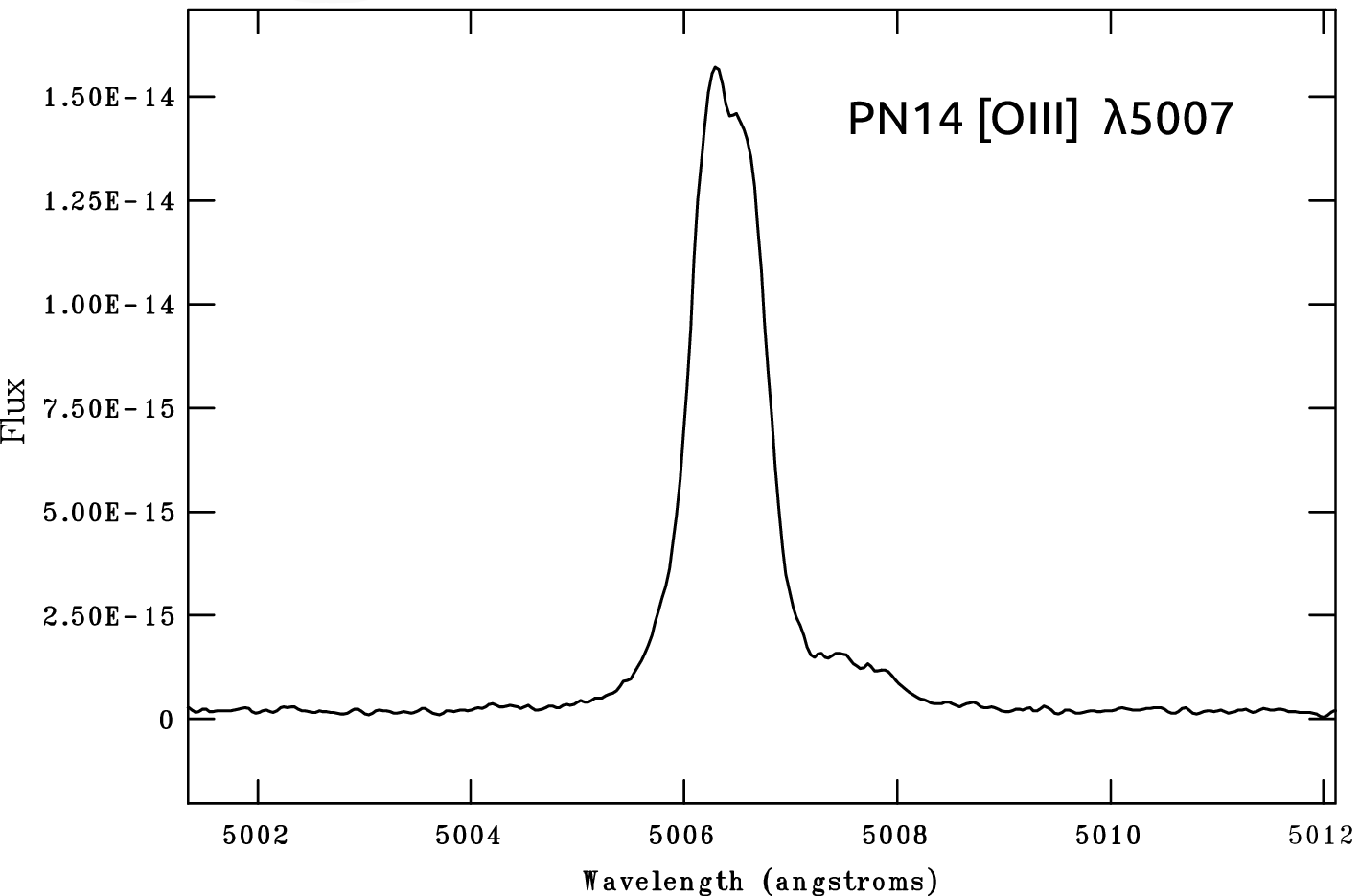}%
     \hfill%
     \includegraphics[width=5.1cm,height=4.45cm]{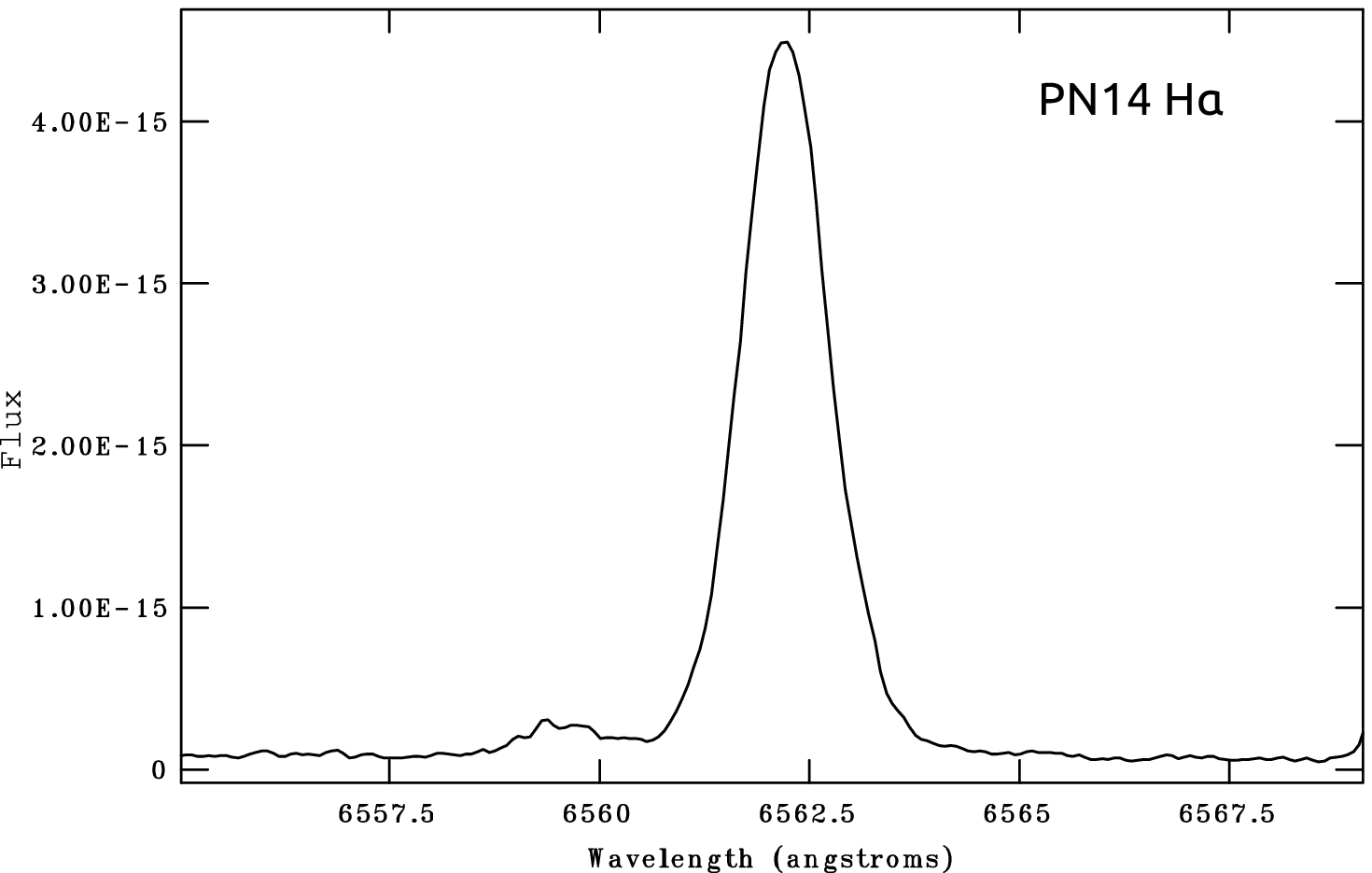}
     \hfill%
	 \includegraphics[width=5.2cm,height=4.8cm]{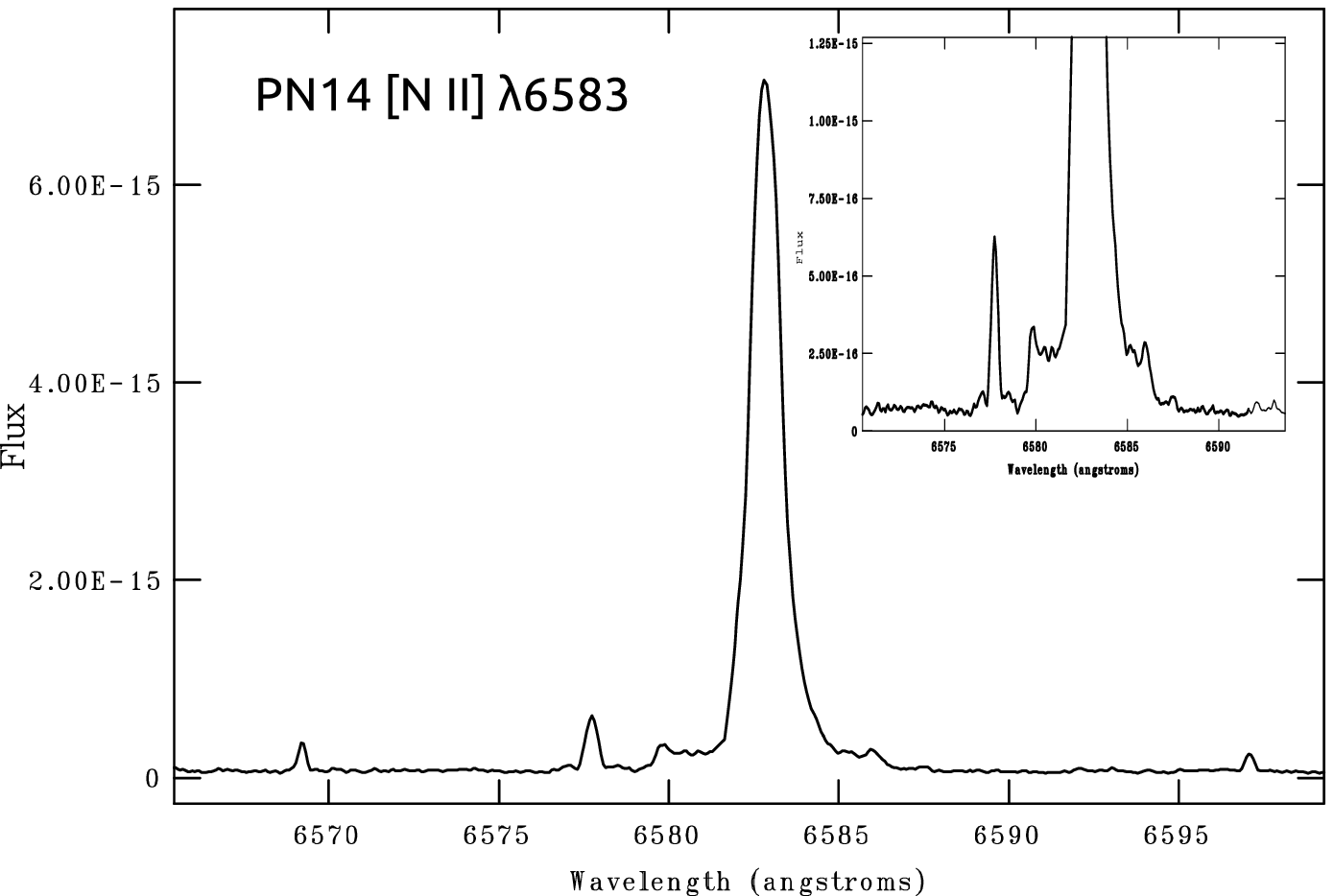}
 	 }	 
  \caption{Line profiles for planetary nebulae. First, second, and third columns show [\ion{O}{iii}] $\lambda$5007, H$\alpha$, and [\ion{N}{ii}] $\lambda$6583 respectively.}
  \label{fig:line-prof}
\end{figure*}

\subsection{Nebular diagnostic}
In our high resolution spectra the auroral line [\ion{O}{iii}] $\lambda$4363 and the [\ion{S}{ii}] $\lambda\lambda$6717,6731, useful  for plasma diagnostic, are measurable for several PNe, allowing us to roughly estimate the nebular electron temperature and density. First, the line fluxes (shown in Table 2) were derreddened. The logarithmic reddening correction at H$\beta$, c(H$\beta$), was obtained from the Balmer decrement. We adopted the interstellar reddening extinction law by Seaton (1979). 

\begin{table*}[!th]\centering
  \caption{Temperature and density in PNe.} 
  \label{tab:temden}
 \begin{tabular}{lcrrr}
    \hline
      & PN6 & PN4 & PN2 & PN14 \\
    \hline
   T(O\,{\sc iii}) (K)	 & 12400 $\pm$ 1900 & 18800 $\pm$ 1630 & 22200 $\pm$ 3900 & 13400 $\pm$ 700 \\
   T(N\,{\sc ii}) (K)   &    ---    &--- &--- & 19500 $\pm$  630\\
   N(S II)(cm$^{-3})$ & 2000 $\pm$ 1400 & 8600 $\pm$ 360 & 480 $\pm$ 290 & 900 $\pm$ 730 \\
   c(H$\beta$) & 0.58 & 0.46 & 0.39 & 0.24 \\
   	\hline
  \end{tabular}
\end{table*}

We were able to determine [\ion{O}{iii}] electron temperatures and [\ion{S}{ii}] densities in PN2, PN4, PN6 and PN14 
although, [\ion{O}{iii}] $\lambda$4363 and [\ion{S}{ii}]$\lambda$6717, 6730  emission lines present large uncertainties. For PN14, the [\ion{N}{ii}] temperature was also  measured through the $\lambda$5755/$\lambda$6583 line ratio. For these calculations  we used the {\it temden} routine implemented in the IRAF NEBULAR package \citep{1995..PASP..107..896S}. The results are presented in Table \ref{tab:temden}. Despite the large errors, the calculated values are similar, whithin uncertainties, to the ones presented by Hern\'andez-Mart{\'\i}nez et al. (2009). Therefore, as PN2 is the only one without previous calculations, we consider its values to be valid . This nebula shows a high electron temperature (about 22000 K), indicating a low metallicity object, therefore it probably belongs to the old objects in the classification of PNe made by  Hern\'andez-Mart{\'\i}nez et al. (2009).

\section{Results}

Most of the results regarding the kinematics are discussed in \S 4, therefore we present a synthesis here. 
From high resolution spectra obtained at LCO and OAN-SPM and data from the literature, the heliocentric radial velocities of ten PNe, four H\,{\sc ii} regions and two A-type supergiant stars were  analyzed and compared to the velocities of the H\,{\sc i} disk at the same position. We found that H\,{\sc ii} regions and A-type supergiants share the kinematics of the rotating H\,{\sc i} disk, while PNe seem to belong to a different kinematical system similar to the one  shown by the C stars. Our result is different from what was found in the irregular galaxy IC\,10, and implies that there are at least two kinematical systems in  NGC\,6822, one defined by the young population (H\,{\sc i} disk, H\,{\sc ii} regions and young stars) the other defined by the intermediate-age population (PNe and C stars), which differ significantly. In addition, four stellar cluster in NGC\,6822 seem to represent an even more different system. These clusters, of low metallicity and at high velocity relative to the system, would have accreted into the halo (Hwang et al. 2014).

Only one object of our sample, named H\,{\sc ii}\,18, has a velocity apparently in contradiction with our results, showing a large velocity difference relative to the H\,{\sc i} gas at the same position.  In the appendix we discuss the possibility that H\,{\sc ii}\,18 is a PN and not an H\,{\sc ii} region, which would explain its kinematics.

Physical conditions of the plasma (electron temperature and density) were determined for some PNe. Our values coincide with previous results from the literature. For the first time, these parameters were derived for PN2, a planetary nebula at the NE of the galaxy, away from the optical bar. The high electron temperature and  radial velocity of PN2 indicate that this is an old  low-metallicity PN.

The line profiles of the PNe observed at LCO were measured. Expansion velocities and some velocity structures were determined for PN2, PN4, PN6, and PN14. The expansion velocities shown by our PNe reach from about  8 to 25 km s$^{-1}$, which is a range similar to the one found in PNe of other galaxies by Richer et al. (2010).  PN14 is a very interesting object. It is a Type I PN (so it is a young object) located near the galactic center, but  it is receding from the galaxy at almost 30 km s$^{-1}$ and  shows a fast bipolar ejection (jet)  with a velocity of about $\pm$140 km s$^{-1}$.

Many objects need to be analyzed in the future to verify the velocity field of the different populations, in particular in the central zone, to constrain the models for computing the distribution of the dark matter halo in NGC\,6822.

\appendix
\section{True nature of the nebula  H\,{\sc  ii}\,18}
\begin{figure}[!th]
  \begin{center}
\includegraphics[width=4.2cm,height=3cm]{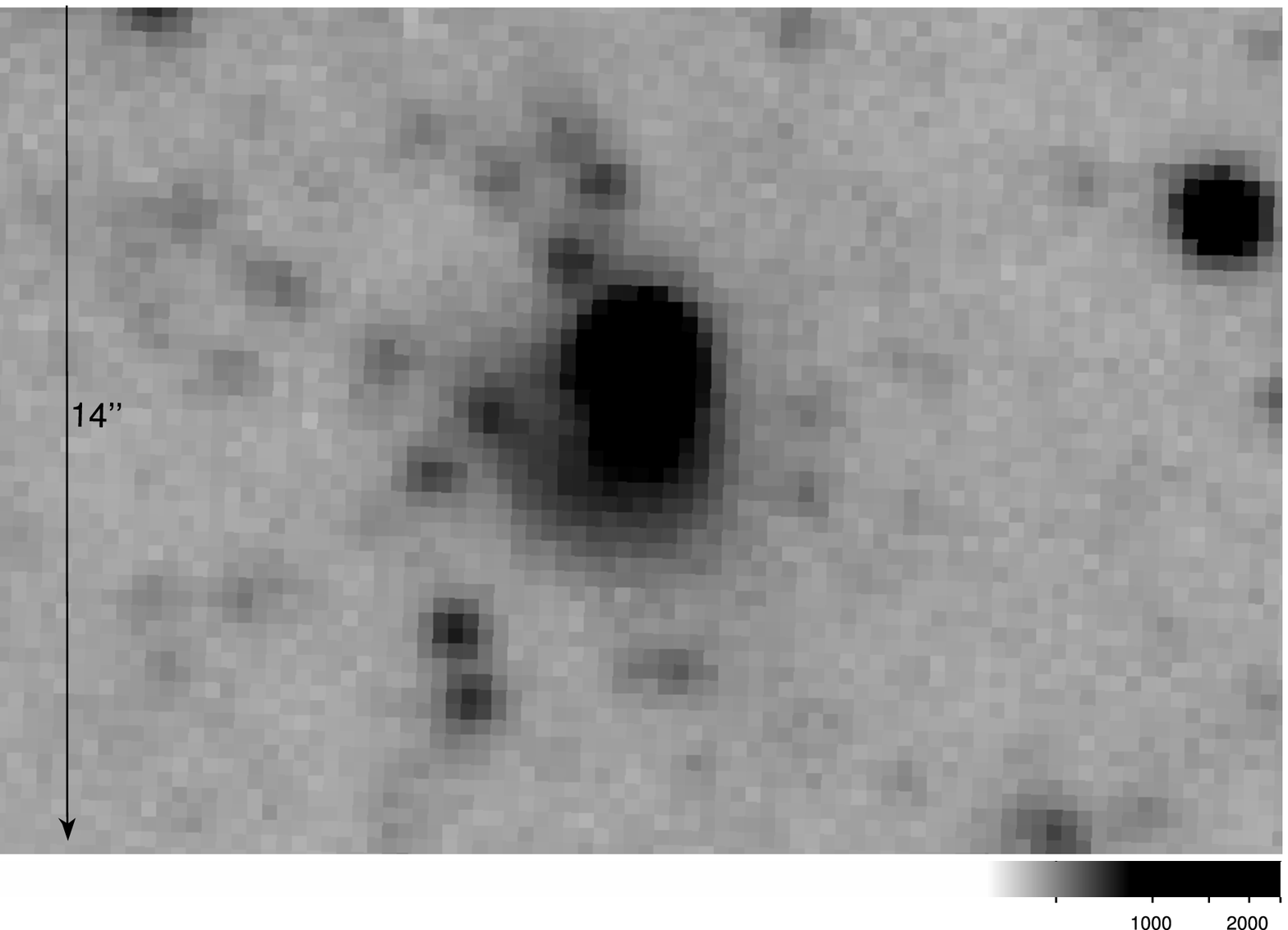}
\includegraphics[width=4.2cm,height=3cm]{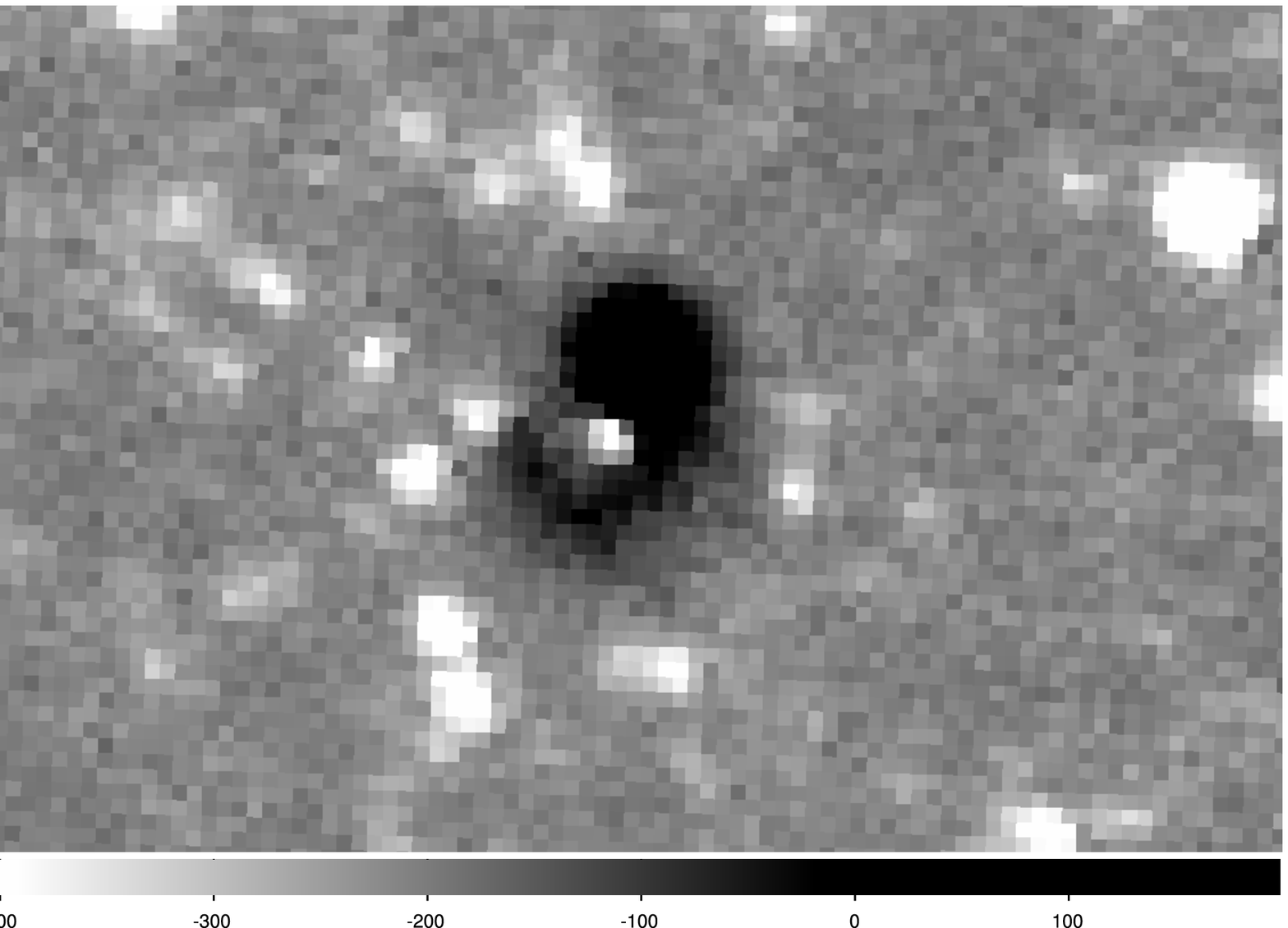}
  \caption{H$\alpha$ (left) and the H$\alpha$-continuum subtracted (right) images of  H\,{\sc  ii}\,18 are shown. North is up and east is to the left. The scale is 14 arcsec. A small faint  H\,{\sc  ii} region is in the SE that seems projected on top of the compact intense H$\alpha$ knot.  \label{fig: HII18}}
\end{center}
\end{figure}
As said in \S3.1, the high radial velocity of this object, compared with that in the  H\,{\sc  i} disk, resembles more the values found for PNe that those of  H\,{\sc  ii} regions. 
This object was first reported by Killen \& Dufour (1982), who named it  S\,10 and described it as a compact nebula. A classification as `PN?' was given to this object. Leisy et al. (2005) declared  it  an extended nebula and classified it as a compact  H\,{\sc  ii} region. It is object number 18 in their list. Similarly, Richer   \& McCall (2007) reported the nebula as extended. These latter authors analyzed spectrophotometric data for this object and, interestingly, they calculated an electron temperature T(\ion{O}{iii}) of almost 15,000 K, which is too high, compared with the values for  H\,{\sc  ii} regions, which never exceed 13,000 K. The oxygen abundance derived by Richer \& McCall is log O/H +12= 7.71, which is very low when compared with the abundances of H\,{\sc ii} regions, which on average are  8.06$\pm$0.04 with a  very small dispersion  (Hern\'andez-Mart{\'\i}nez et al. 2009). With this high temperature and low metallicity,  H\,{\sc  ii}\,18 is more similar to the PNe of low chemical abundances in this galaxy. 

Another characteristic of  H\,{\sc  ii}\,18 that might be indicating  a PN nature rather than a H\,{\sc ii} region is  its  H$\alpha$ luminosity. By considering the distance given by Gieren et al. (2006) and the H$\alpha$ flux estimated by Hern\'andez-Mart{\'\i}nez \& Pe\~na (2009), we obtain  L(H$\alpha$)/L$_\odot$ = 85.5, which is very similar to the H$\alpha$ luminosity of PN14 (76.9 L$_\odot$) and is lower than expected for a compact H\,{\sc ii} region ionized by an OB star.

 Considering all the above, we revised the images in [\ion{O}{iii}]$\lambda$ 5007 and H$\alpha$ obtained with the VLT FORS2 spectrograph on 2006,  which served as pre-imaging  for the spectroscopy work (program ID 077.B-0430) reported by Hern\'andez-Mart{\'\i}nez et al. (2009). The H$\alpha$ image and the subtracted H$\alpha$-continuum image are shown in Fig.A.1  H\,{\sc  ii}\,18 appears as a very intense compact nebula located at the edge of a faint shell-like  H\,{\sc  ii} region , whose central star seems clearly visible. 

Based on this evidence, we suggest that  H\,{\sc  ii}\,18 is a planetary nebula, overlapping a  H\,{\sc  ii} region, which gives it appearance of extended. However, our discussion is not conclusive. Deep spectroscopic data and highly resolved imaging would be required to definitely establish the true nature of the H\,{\sc ii}18 - S10 nebula.

\begin{acknowledgements}
We are indebted to W. J. G. de Blok and Antonio Peimbert, who kindly provided the  H\,{\sc  i} velocity map and the spectra of H\,V and H\,X used in this work. Enlightening discussion with Michael Richer is deeply appreciated. S.F.-D. received scholarship from  CONACyT-Mexico, and from DGAPA-PAPIIT (UNAM) grant IN105511. This work received financial support from DGAPA-PAPIIT (UNAM) grants IN105511 and 109614. 
\end{acknowledgements}


\begin{thebibliography}{}	

\bibitem[{{Akras} \& {L\'opez}(2012)}]{akras} {Akras}, S. \& {L\'opez}, J. A. 2012, MNRAS, 425, 2197 
\bibitem[{{Allen} {et~al.}(1998)}]{allen} {Allen}, C., {Carigi}, L., \& {Peimbert}, M.  1998, ApJ, 494, 247
\bibitem[{{Batinelli} {et~al.}(2006)}]{batinelli} {Battinelli},  P., {Demers}, S., \& {Kunkel}, W. E.  2006, A\&A, 451, 905
\bibitem[{{Berstein} {et al.}(2003)}]{berstein} {Berstein},  R. A.,  {Shectman}, S. A., {Gunnels}, S.,  {Mochnacki}, S., \& {Athey}, A. 2002, Proc. SPIE 4841
\bibitem[{{Coccato} {et~al}(2009)}]{coccato}  Coccato, L., Gerhard, O., Arnaboldi, M., Das, P.,  et al. 2009, MNRAS, 394, 1249
\bibitem[{{de Blok} \& {Walter}(2000)}]{2000ApJ..537..L95B} {de Blok}, W.~J.~G, \& {Walter}, F.\ 2000, ApJ, 537, L95 
\bibitem[{{de Blok} \& {Walter}(2006)}]{2006AJ..131..343B} {de Blok}, W.~J.~G, \& {Walter}, F.\ 2006, AJ, 131, 343
\bibitem[{{Demers} {et~al.}(2006)}]{2006ApJ..636..L85D} {Demers}, S., {Battinelli}, P., \& {Kunkel}, W.~E.\ 2006, ApJ, 636, L85 
\bibitem[{{Garc\'{\i}a-Rojas} {et~al.}(2012)}]{Garcia-Rojas et al} {Garc\'{\i}a-Rojas}, J., {Pe\~na}, M., {Morisset}, C., {Mesa-Delgado}, A., \& {Ruiz}, M. T. 2012, A\&A, 538, A54
\bibitem[{{Gerhard} {et al.}(2007)}]{gerhard et al.}{Gerhard}, O., {Arnaboldi}, M., {Freeman}, K. C., et al. 2007,  A\&A, 468, 815
\bibitem[{{Gieren} {et~al.}(2006)}]{gieren} {Gieren}, W.,  {Pietrzy\'nski}, G., {Nalewajko}, K., et al. 2006, ApJ, 647, 1056
\bibitem[{{Gon\c{c}alves} {et~al.}(2012)}]{} {Gon\c{c}alves}, D.~R., {Teodorescu}, A.~M., {Alves-Brito}, A., {M\'endez}, R.~H., \& {Magrini}, L. 2012, MNRAS, 425, 2557
\bibitem[{{Hern\'andez-Mart\'inez} \& {Pe\~na}(2009)}]{2009A&A..495..447} {Hern\'andez-Mart\'inez}, L., \& {Pe\~na}, M. \ 2009, A\&A, 495, 447 
\bibitem[{{Hern\'andez-Mart\'inez} {et~al.}(2009)}]{2009A&A..436..437} {Hern\'andez-Mart\'inez}, L., {Pe\~na}, M., {Carigi}, L., \& {Garc\'ia-Rojas}, J. \ 2009, A\&A, 505, 1027
\bibitem[{{Hodge}(1977)}]{1977ApJS..33..69H} {Hodge}, P.~W. \ 1977, ApJS, 33, 69  
\bibitem[{{Hodge} {et~al.}(1988)}]{Hodge etal.} {Hodge}, P. W., {Lee}, M.~G. \& {Kennicutt}, R.~C.~Jr. 1988, PASP, 100, 917
\bibitem[{{Hubble} (1925)}]{hubble} {Hubble}, E. 1925, ApJ, 62, 409 
\bibitem[{{Hwang} {et al.}(2014)}] {Hwang et al.} {Hwang}, N., Park, H. S.,  Lee, M. G., et al., 2014, ApJ, 783, 49
\bibitem[{{Killen} \& {Dufour}(1982)}]{1982PASP.94.444}{Killen}, R. M., \& {Dufour}, R.J. 1982, PASP, 94, 444
\bibitem[{{Leisy} {et~al.}(2005)}]{2005A&A..436..437L} {Leisy}, P., {Corradi}, R.~L.~M., {Magrini}, L., {Greimel}, R., {Mampaso}, A., \& {Dennefeld}, M.\ 2005, A\&A, 436, 437 
\bibitem[{{Letarte} {et~al.}(2002)}]{2002AJ..123..832L} {Letarte}, B., {Demers}, S., {Battinelli}, P., \& {Kunkel}, W.~E.\ 2002, AJ, 123, 832 
\bibitem[{{Mateo}(1998)}]{1998ARA&A..36..435M} {Mateo}, M. \ 1998, ARA\&A, 36, 435
\bibitem[{{McNeil}(2012)}] {2012AA539.11} {McNeil-Moylan}, E. K., Freeman, K. C., Arnaboldi, M., \& Gerhard, O. E.  \ 2012, A\&A, 539, 11
\bibitem[{{Medina}{et~al.}(2006)}]{Medina et al.} {Medina}, S., {Pe\~na}, M., {Morisset}, C. \& {Stasi\'nska}, G. 2006, RevMexAA, 42, 53
\bibitem[{{Navarro}(1996)}]{1996ApJ..462..563}{Navarro}, J. F., {Frenk}, C. S., \& {White}, S. D. M. 1996, ApJ, 462, 563 
\bibitem[{{Oke}(1990)}]{Oke} {Oke},  J. B. 1990, AJ, 99, 1621
\bibitem[{{Richer} \& {McCall}(2007)}]{2007ApJ..658..328}{Richer}, M.~G., {McCall}, M.~L. 2007, ApJ, 658, 328
\bibitem[{{Richer} {et~al.}(2010)}]{2010RMxAA..46..191R} {Richer}, M.~G., {L\'opez}, J.~A., {D\'iaz-M\'endez}, E., et al.\ 2010, RevMexAA, 46, 191
\bibitem[{{Peimbert} {et~al.}(2005)}]{2005ApJ..634..1056P}{Peimbert}, A., {Peimbert}, M. \& {Ru\'iz}, Ma.~T.\ 2005, ApJ, 634, 1056P
\bibitem[{{Seaton}(1979)}]{1979..MNRAS..185..5S}{Seaton}, M. \ 1979, MNRAS, 185, 5
\bibitem[{{Shaw} \& {Dufour}(1995)}]{1995..PASP..107..896S} {Shaw}, R.~A. \& {Dufour}, R.~J.\ 1995, PASP, 107, 896
\bibitem[{{Teodorescu} {et al.}(2011)}]{2011..ApJ..736..65}  {Teodorescu}, A. M., {M\'endez}, R H., {Bernardi}, F., {Thomas}, J., {Das}, P, \& Gerhard, O. \ 2011, ApJ, 736, 65
\bibitem[{{Valenzuela} {et~al.}(2007)}]{2007ApJ..657..773V}{Valenzuela}, O., {Rhee}, G., {Kyplin}, A., {Governato}, F., {Stinson}, G. {Quinn}, T., \& {Wadsley}, J.\ 2007, ApJ, 657, 773 
\bibitem[{{Venn} {et~al.}(2001)}]{venn}{Venn}, K. A. , {Lennon}, D. J., {Kaufner}, A., et al.\ 2001, ApJ, 547, 765
\bibitem[{{Ventimiglia} {etal.} (2011)}]{ventimiglia} {Ventimiglia}, G., {Arnaboldi}, M., \& {Gerhard}, O. 2011, A\&A, 528, 24
\bibitem[{{Weldrake} {et~al.}(2003)}]{2003MNRAS..340..12W}{Weldrake}, D.~T.~F., {de Blok}, W.~J.~G., \& {Walter}, F.\ 2003, MNRAS, 340, 12
\end{thebibliography}
\end{document}